%% file: manuscript.tex
\documentclass[sigconf, screen]{acmart}
\acmConference[ESEC/FSE 2021]{The 29th ACM Joint European Software Engineering Conference and Symposium on the Foundations of Software Engineering}{23 - 27 August, 2021}{Athens, Greece}
\renewcommand\footnotetextcopyrightpermission[1]{}
\acmDOI{10.0000/00.000}
\usepackage{bookmark}

\usepackage[inline]{enumitem}
\usepackage{fontawesome}
\usepackage{colortbl}
\usepackage{wrapfig}
\usepackage{graphicx}

\usepackage{footnote}
\makesavenoteenv{tabular}

\usepackage{amsmath}
\usepackage{amsfonts}

\usepackage{algorithmicx}
\usepackage{multirow}
\usepackage{multicol}
\usepackage{booktabs}

\usepackage{array}
\usepackage{xspace}
\usepackage{bigstrut}
\usepackage[export]{adjustbox}

\usepackage[caption=false,font=footnotesize]{subfig}

\usepackage{stfloats}

\usepackage{boldline}

\usepackage{cleveref}
\crefformat{section}{\S#2#1#3}
\crefmultiformat{section}{\S\S#2#1#3}{and~#2#1#3}{, #2#1#3}{, and~#2#1#3}
\Crefname{equation}{Eq.}{Eqs.}
\Crefname{figure}{Fig.}{Figs.}
\Crefname{tabular}{Tab.}{Tabs.}
\crefname{algocf}{pseudocode}{pseudocodes}
\Crefname{algocf}{Pseudocode}{Pseudocodes}

\input{pygments}

\input{macros}

\input{listings.tex}
\usepackage{url}

\usepackage{titlesec}

\acmConference[Preprint]{}{October, 2020}{ARXIV}

\author{Md Shahriar Iqbal}
\authornote{Joint First Author}
\affiliation{%
  \institution{University of South Carolina}}
\email{miqbal@email.sc.edu}

\author{Rahul Krishna}
\authornotemark[1]
\affiliation{%
  \institution{Columbia University}}
\email{rahul.krishna@columbia.edu}

\author{Mohammad Ali Javidian}
\affiliation{%
  \institution{Purdue University}}
\email{mjavidia@purdue.edu}

\author{Baishakhi Ray}
\affiliation{%
  \institution{Columbia University}}
\email{rayb@cs.columbia.edu}

\author{Pooyan Jamshidi}
\affiliation{%
  \institution{University of South Carolina}}
 \email{pjamshid@cse.sc.edu}

\begin{document}
\title{\tool: Debugging and Fixing Misconfigurations\\ using Counterfactual Reasoning}

\input{0_abstract.tex}
\maketitle
\input{1_intro.tex}
\input{2_motivate.tex}
\input{3_formulation.tex}

\input{4_method.tex}

\input{11_case_study.tex}
\input{5_evaluation.tex}
\input{6_results.tex}

\input{9_related.tex}

\input{10_conclusion.tex}
\balance
\bibliographystyle{ACM-Reference-Format}
\bibliography{references,bib_pjamshidi}
\end{document}

%% file: pygments.tex
\definecolor{gray05}{gray}{0.95}
\definecolor{gray10}{gray}{0.90}
\definecolor{gray12}{gray}{0.88}
\definecolor{gray15}{gray}{0.85}
\definecolor{gray20}{gray}{0.80}
\definecolor{gray25}{gray}{0.75}
\definecolor{gray30}{gray}{0.70}
\definecolor{gray40}{gray}{0.60}
\definecolor{gray50}{gray}{0.50}
\definecolor{gray60}{gray}{0.40}
\definecolor{gray70}{gray}{0.30}
\definecolor{gray75}{gray}{0.25}
\definecolor{gray80}{gray}{0.20}
\definecolor{gray85}{gray}{0.15}
\definecolor{gray90}{gray}{0.10}
\definecolor{gray95}{gray}{0.05}

\newcommand\red[1]{\textcolor[RGB]{237,28,36}{#1}}

%% file: macros.tex
\newcommand\eq[1]{\Cref{eq:#1}\xspace}
\newcommand\fig[1]{\Cref{fig:#1}\xspace}
\newcommand\tab[1]{\Cref{tab:#1}\xspace}
\newcommand\tion[1]{\Cref{sect:#1}\xspace}

\newcommand{\etal}{\textit{et al.}\xspace}
\newcommand{\ie}{i.e.}
\newcommand{\eg}{e.g.}

\newcommand{\etc}{\textit{etc}\xspace}

\newcommand{\tool}{\textsc{CADET}\xspace}
\newcommand{\dd}{$\delta$\textsc{-Debug~}\xspace}
\newcommand{\encore}{\textsc{EnCore}\xspace}
\newcommand{\bugdoc}{\textsc{BugDoc}\xspace}
\newcommand{\cbi}{\textsc{CBI}\xspace}

\newcommand\txone{\textsc{TX1}\xspace}
\newcommand\txtwo{\textsc{TX2}\xspace}
\newcommand\xavier{\textsc{Xavier}\xspace}

\newcommand\gpugrowth{\texttt{GPU} \texttt{memory} \texttt{growth}\xspace}
\newcommand\swapmem{\texttt{swap} \texttt{memory}\xspace}

\newcommand{\nfp}{\textsc{nf-fault}\xspace}
\newcommand{\nfps}{\textsc{nf-faults}\xspace}

\newcommand{\be}{\begin{enumerate}}
\newcommand{\smalleq}{\begin{equation}\small}
\newcommand{\smallereq}{\begin{equation}\footnotesize}
\newcommand{\eeq}{\end{equation}}
\newcommand{\beqml}{\begin{multline}}
\newcommand{\eeqml}{\end{multline}}
\newcommand{\besq}{\begin{enumerate}[leftmargin=*,wide=0pt,topsep=0pt]}
\newcommand{\ee}{\end{enumerate}}
\newcommand{\bi}{\begin{itemize}}
\newcommand{\bicirc}{\begin{itemize}[leftmargin=*]\renewcommand\labelitemi{$\circ$}}
\newcommand{\bisq}{\begin{itemize}[leftmargin=*,wide=0pt,topsep=0pt]}
\newcommand{\ei}{\end{itemize}}

\makeatletter
\edef\textFontName{\fontname\csname
  \f@encoding/\f@family/\f@series/\f@shape/\f@size\endcsname}

\newcommand{\removelatexerror}{\let\@latex@error\@gobble}

\makeatother

\usepackage{tikz}
\usetikzlibrary{arrows.meta}

\newcommand\edgeone{
\begin{tikzpicture}
  \draw[black,  arrows={-Triangle[angle=90:3pt,black,fill=black]}] (0,0.0) -- (0.5,0.0);
\end{tikzpicture}\xspace}

\newcommand\edgetwo{
\begin{tikzpicture}
  \draw[black,  arrows={Triangle[angle=90:3pt,black,fill=black]-Triangle[angle=90:3pt,black,fill=black]}] (0,0.0) -- (0.5,0.0);   
\end{tikzpicture}\xspace}

\newcommand\edgethree{
\begin{tikzpicture}
  \draw[black,  arrows={Circle[open]-Triangle[angle=90:3pt,black,fill=black]}] (0,0.0) -- (0.5,0.0);   
\end{tikzpicture}\xspace}

\newcommand\edgefour{
\begin{tikzpicture}
  \draw[black,  arrows={Circle[open]-Circle[open]}] (0,0.0) -- (0.5,0.0);   
\end{tikzpicture}\xspace}

\usepackage{amsthm}
\DeclareMathOperator*{\argmax}{argmax}
\usepackage{thmtools}
\usepackage{tikz}
\usetikzlibrary{arrows.meta}
\declaretheoremstyle[headfont=\scshape]{schead}
\declaretheoremstyle[headfont=\bf]{bfhead}

\usepackage[linesnumbered,ruled,vlined, noend]{algorithm2e}

\SetCommentSty{mycommfont}
\SetAlCapNameFnt{\footnotesize}
\SetAlCapFnt{\footnotesize}
\SetAlgorithmName{Pseudocode}{Pseudocode}{Pseudocode of Algorithms}

\usepackage[framemethod=tikz]{mdframed}
\usetikzlibrary{shadows}
\usepackage{graphics}
\newmdenv[
    tikzsetting= {fill=gray05!0},
    skipabove=0.33em,
    skipbelow=0.33em,
    linewidth=1pt,
    innerleftmargin=4pt,
    innerrightmargin=4pt,
    innertopmargin=2pt,
    innerbottommargin=2pt,
    linecolor=gray85,
    roundcorner=2pt, 
    shadow=true,
    shadowsize=4pt,
    shadowcolor=black
]{myshadowbox}
\newmdenv[
    tikzsetting= {fill=gray05!0},
    skipabove=0.33em,
    skipbelow=0.33em,
    linewidth=1.25pt,
    innerleftmargin=4pt,
    innerrightmargin=4pt,
    innertopmargin=2pt,
    innerbottommargin=2pt,
    linecolor=gray85,
    roundcorner=3pt, 
    shadow=false,
    shadowsize=3pt,
    shadowcolor=black
]{mygoalbox}
\usepackage{tikz}


%% file: 0_abstract.tex
\begin{abstract}

%
Modern computing platforms are highly-configurable with thousands of interacting configuration options. However, configuring these systems is challenging and misconfigurations can cause unexpected non-functional faults. 
%
%
This paper proposes \tool (short for \underline{Ca}usal \underline{De}bugging \underline{T}oolkit) that enables users to \emph{identify, explain}, and \emph{fix} the root cause of non-functional faults early and in a principled fashion. 
%
%
\tool builds a causal model by observing the performance of the system under different configurations. Then, it uses casual path extraction followed by counterfactual reasoning over the causal model to (a)~identify the root causes of non-functional faults, (b)~estimate the effects of various configuration options on the performance objective(s), and (c)~prescribe candidate repairs to the relevant configuration options to fix the non-functional fault. 
%
%
We evaluated \tool on 5 highly-configurable systems by comparing with state-of-the-art configuration optimization and ML-based debugging approaches. The experimental results indicate that \tool can find effective repairs for faults in multiple non-functional properties with (at most) 13\% more accuracy, 32\% higher gain, and $13\times$ speed-up than other ML-based performance debugging methods. Compared to multi-objective optimization approaches, \tool can find fixes (at most) $8\times$ faster with comparable or better performance gain. Our study of non-functional faults reported in NVIDIA's forum shows that \tool can find $14\%$ better repairs than the experts' advice in less than 30 minutes.




\end{abstract}
    

%% file: 1_intro.tex
\section{Introduction}

Modern computing systems are highly configurable and can seamlessly be deployed on various hardware platforms and under different environmental settings. The configuration space is combinatorially large with 100s if not 1000s of software and hardware configuration options that interact non-trivially with one another~\cite{wang2018understanding, halin2019test,JC:MASCOTS16}. Unfortunately, configuring these systems to achieve specific goals is challenging and error-prone.

Incorrect configuration (\textit{misconfiguration}) elicits unexpected interactions between software and hardware resulting \textit{non-functional faults}, \ie, faults in \textit{non-functional} system properties such as latency and energy consumption. These non-functional faults---unlike regular software bugs---do not cause the system to crash or exhibit an obvious misbehavior~\cite{reddy2016fault, tsakiltsidis2016automatic, nistor2013discovering}. Instead, misconfigured systems remain operational while being compromised, resulting severe performance degradation in latency, energy consumption, and/or heat dissipation~\cite{bryant2003computer, molyneaux2009art,sanchez2020tandem,nistor2015caramel}. 
The sheer number of modalities of software deployment is so large that exhaustively testing every conceivable software and hardware configuration is impossible. 
Consequently, identifying the root cause of non-functional faults is notoriously difficult~\cite{gunawi2018fail} with as much as 99\% of them going unnoticed or unreported for extended durations~\cite{99ofmisc75:online}. This has tremendous monetary repercussions costing companies worldwide an estimated \$5 trillion in 2018 and 2019~\cite{Cloudmis78:online}. Further, developers on online forums are quite vocal in expressing their dissatisfaction. For example, one developer on NVIDIA's developer forum bemoans: \textit{``I am quite upset with CPU usage on \txtwo~\cite{HighCPUu7:online},''} while another complained, \textit{``I don’t think it [the performance] is normal and it gets more and more frustrating~\cite{super_frustrated_power_perf:online}.''} Crucially, these exchanges provoke other unanswered questions, such as,~``\textit{what would be the effect of changing another configuration `X'?~\cite{slowimag79:online}}.'' Therefore, we seek methods that can \emph{identify, explain, and fix} the root cause of non-functional faults early in a principled fashion.

\input{figures/software_stack.tex}

\noindent\textbf{Existing work.}~
Much recent work has focused on configuration optimization, which are approaches aimed at finding a configuration that optimizes a performance objective~\cite{wu2015deep, yigitbasi2013towards,siegmund2015performance, nair2018faster, filieri2015automated}. Finding the optimum configuration using push-button optimization approaches is not applicable here because they do not give us any information about the underlying interactions between the faulty configuration options that caused the non-functional fault. This information is sought after by developers seeking to address these non-functional faults~\cite{reddy2016fault, thung2012empirical}. 

Some previous work has used machine learning-based performance modeling approaches~\cite{guo2013variability, siegmund2015performance, valov2017transferring, siegmund2012predicting}. These approaches are adept at inferring the correlations between certain configuration options and performance, however, they lack the mathematical language to express \textit{how} or \textit{why} the configuration options affect performance. Without this, we risk drawing misleading conclusions. They also require significant time to gather the training samples, which grows exponentially with the number of configurations~\cite{kanellis2020too,xu2015hey}.  


\noindent\textbf{Limitations of existing work.}~
In \fig{intro_fig}, we present an example to help illustrate the limitations with current techniques. Here, the observational data gathered so far indicates that \gpugrowth is positively correlated with increased latency (as in Fig.~\ref{fig:intro_fig}a). An ML-model built on this data will, with high confidence, predict that larger \gpugrowth leads to larger latency. However, this is counter-intuitive because higher \gpugrowth should, in theory, reduce latency not increase it. When we segregate the same data on \swapmem (as in~Fig.~\ref{fig:intro_fig}b), we see that there is indeed a general trend downward for latency, \ie, within each group of \swapmem, as \gpugrowth increases the latency decreases.
We expect this because \gpugrowth controls how much memory the GPU can ``borrow'' from the \swapmem. Depending on resource pressure imposed by other host processes, a resource manager may arbitrarily re-allocate some \swapmem; this means the GPU borrows proportionately more/less swap memory thereby affecting the latency correspondingly. This is reflected by the data in Fig.~\ref{fig:intro_fig}b. If the ML-based model were to consult the available data (from Fig.~\ref{fig:intro_fig}a) unaware of such underlying factors, these models would be incorrect.
With thousands of configurations, inferring such nuanced information from optimization or ML-based approaches would require a considerable amount of measurements and extensive domain expertise which can be impractical, if not impossible, to possess in practice.

%



\noindent\textbf{Our approach.}~
In this paper, we design, implement, and evaluate \tool (short for \underline{Ca}usal \underline{De}bugging \underline{T}oolkit). \tool uses causal structural discovery algorithms~\cite{spirtes2000causation} to construct a causal model using observational data~\cite{pearl2009causality}. Then, it uses counterfactual reasoning over the causal model to (a)~identify the root causes of non-functional faults, (b)~estimate the effects of various configurable parameters on the non-functional properties, and (c)~prescribe candidate repairs to the relevant configuration options to fix the non-functional fault. 
%
%
%
%
For example, in~\fig{intro_fig}, \tool constructs a \textit{causal model} from observational data (as in Fig.~\ref{fig:intro_fig}c). This causal model indicates that \gpugrowth indirectly influences latency (via a \swapmem) and that the configuration options may be affected by certain other factors, \eg, resource manager allocating resources for other processes running on the host. \tool uses counterfactual questions such as, ``what is the effect of \gpugrowth on latency if the available \swapmem was 2Gb?'' to diagnose the faults and recommend changes to the configuration options to mitigate these faults.

\noindent\textbf{Evaluation.}~
We evaluate \tool on 5 real-world highly configurable systems (three machine learning systems, a video encoder, and a database system) deployed on 3 architecturally different NVIDIA Jetson systems-on-chip. We compare \tool with state-of-the-art configuration optimization and ML-based performance debugging approaches. Overall, we find that \tool is (at most) $13\times$ faster with $13\%$ better accuracy and $32\%$ higher gain than the next best ML-based approaches for single-objective faults. Compared to single-objective optimization approaches, \tool can find repairs to misconfigurations $9\times$ faster while performing as well as (or better than) the best configuration discovered by the optimization techniques. Further, \tool can find effective repairs for faults in multiple performance objectives (\ie, latency and energy consumption) with accuracy and gain as high as $16\%$ and $36\%$, respectively, than other ML-based performance debugging methods. Compared to multi-objective optimization approaches, \tool can find repairs $5\times$ faster while having similar performance gain. Finally, with a case study, we demonstrate that \tool finds $14\%$ better repairs than the experts' advice in 24 minutes.

\noindent\textbf{Contributions.}~In summary, our contributions are as follows.
\bisq
    \item We propose \tool, a novel causal diagnostics, and fault mitigation tool that identifies the root causes of non-functional faults and recommends repairs to resolve non-functional faults.
    \item Our empirical results, conducted on on 5 highly configurable systems deployed on 3 architecturally different hardware, \tool outperforms current state-of-the-art ML-based diagnostics and optimization approaches in terms of efficacy (+16\% accuracy and +36\% gain) and speed (up to 13$\times$ faster). \tool was also adept at handling both single- and multi-objective performance faults. 
    \item We offer a manually curated performance fault dataset (called Jetson Faults) and accompanying code required to reproduce our findings at \url{https://git.io/JtFNG}. 
\ei    

%% file: figures/software_stack.tex
\begin{figure}[t]
    \setlength{\belowcaptionskip}{-1.25em}
    \centering
    \includegraphics[width=\linewidth]{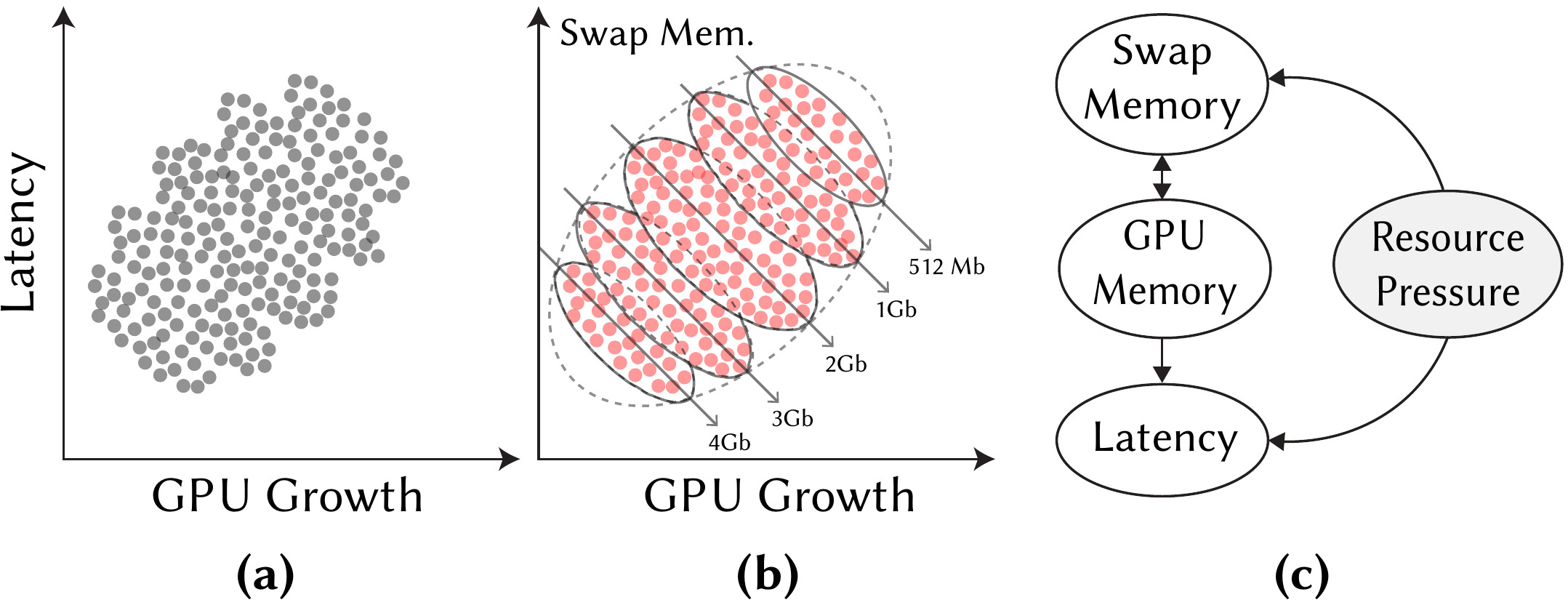}
    \caption{Observational data (in Fig.~\protect\ref{fig:intro_fig}a)  (incorrectly) shows that high \gpugrowth leads to high latency. The trend is reversed when the data is segregated by \swapmem.}
    \label{fig:intro_fig}
\end{figure}


%% file: 2_motivate.tex
\section{Motivation}
\label{sect:motivation}

\subsection{A real-world example}
\label{sect:motivating_examples}

This section illustrates the challenges both users and component developers face
when configuring and composing complex systems with a concrete issue
report from the NVIDIA forum~\cite{code_transplant:online}. Here, a user notices some strange behavior when trying to transplant their code for real-time computation of depth information from stereo-cameras from NVIDIA Jetson \txone to \txtwo. 
Since \txtwo has twice the computational power, they expected at least 30\% lower latency, but observed 4 times higher latency. The user indicates working with a reliable reference implementation of their code and is puzzled by the occurrence of this fault. 

To solve this, the user solicits advice from the NVIDIA forums. After discussions spanning two days, they learn that they have made several misconfigurations:
\begin{enumerate*}
\item \textbf{Wrong compilation flags}:~Their compilation does not take into account the microarchitectural differences between the two platforms. These may be fixed by setting the correct microarchitecture with \texttt{-gencode=arch} parameter and compiling the code dynamically by disabling the \texttt{CUDA\_USE\_STATIC} flag. 
\item \textbf{Wrong CPU/GPU clock frequency}: Their hardware configuration is set incorrectly. These may be fixed by setting the configuration \texttt{-nvpmodel=MAX-N} which changes the CPU and GPU clock settings. The \texttt{Max-N} setting in \txtwo provides almost twice the performance of TX1~\cite{tx1_v_ts2:online} due to a number of factors including increased clock speeds and \txtwo's use of 128-bit memory bus width versus the 64-bit in TX1~\cite{tx1_v_ts2:online}. If \txtwo is not configured to leverage these differences, it will face high latency. 
\item \textbf{Wrong fan modes}: Their fan needs to be configured correctly to achieve higher CPU/GPU clock speeds. If the fan modes are not configured to be high, \txtwo will thermal throttle the CPU and GPU to prevent overheating~\cite{thermal_management:online} and invariably increasing  latency~\cite{chiu2012method}. 
\end{enumerate*}

\subsection{Non-functional faults (\nfps)}
\begin{wrapfigure}[16]{R}{0.45\linewidth}
    \centering
    \includegraphics[width=\linewidth]{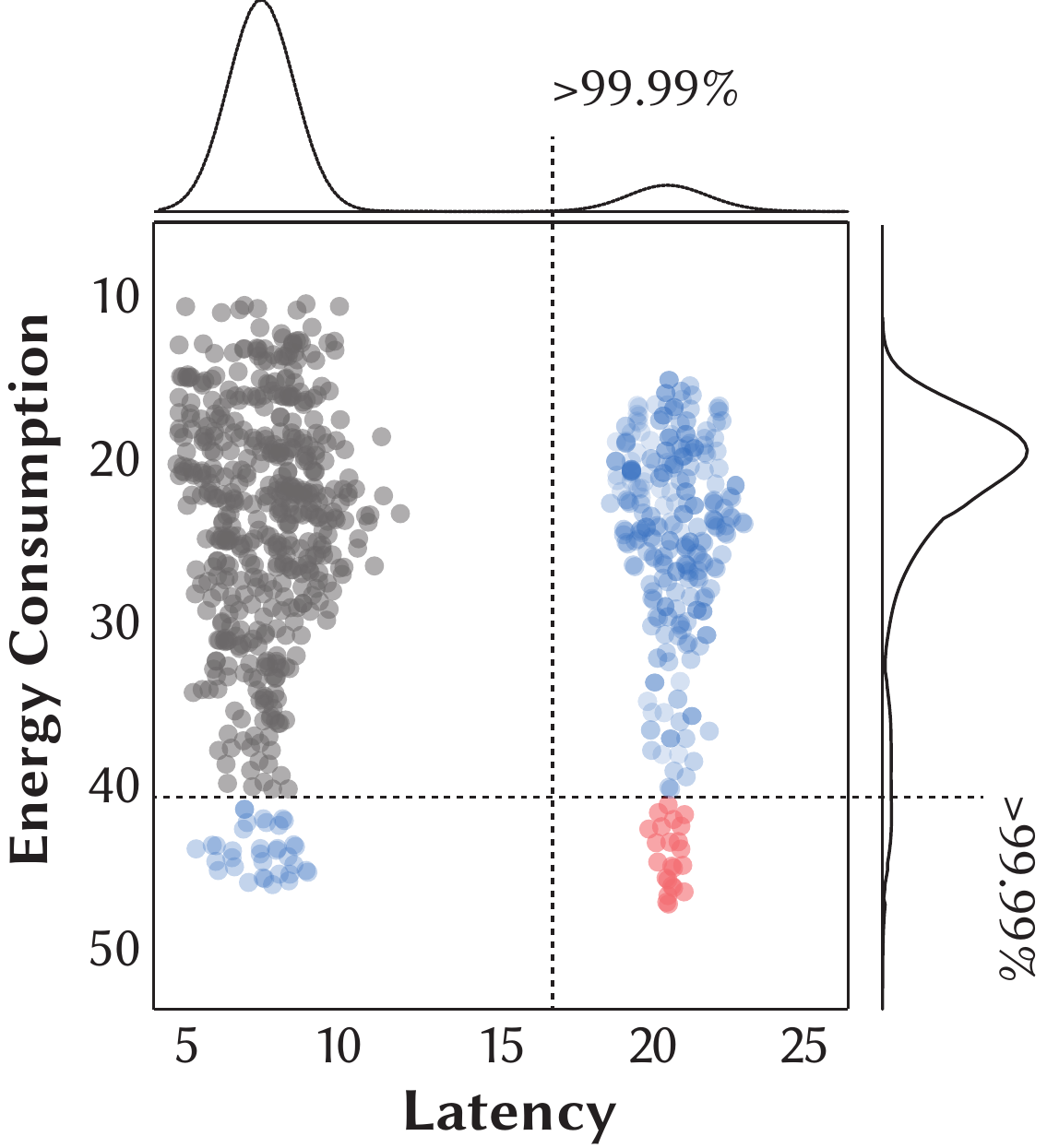}
    \caption{\small An example of multi-objective \nfps.}
    \label{fig:faults_eg}
\end{wrapfigure}
The previous example is one of many examples wherein misconfigurations severely impact performance and energy consumption. Such examples abound in many other systems and domains, including IoT ubiquitous systems (e.g. Amazon Alexa)~\cite{Keepingt50:online,AlexaHac23:online} and production-scale cloud-based systems~\cite{gunawi2018fail}.
We term such anomalies as non-functional faults (hereafter, \nfps), and they represent a mismatch between the expected performance of a non-functional property (such as the latency or energy consumption) and what is actually observed when a system is deployed. Specifically, for certain configurations the software system may experience \textit{tail values} of latency, energy consumption, or heat dissipation, \ie, they take values worse than predetermined SLA (we choose performance worse than 99th percentile). In certain cases, the software system might suffer from tail values in multiple non-functional properties simultaneously, \ie, multi-objective \nfps. 

\Cref{fig:faults_eg} illustrates the distribution of two non-functional properties (latency and energy consumption) measured for 2000 different configuration setting for an NVIDIA TX2 SoC running an image recognition task. There exist two types of non-functional faults that may be discerned from~\Cref{fig:faults_eg}: \textit{(1)} configurations whose latency \textit{or} energy is worse that $99.99\%$ of other configurations (shown in {\color[RGB]{106,166,215} $\bullet$}); \textit{(2)} configurations where \textit{both latency and energy} is worse that $99.99\%$ of other configurations (shown in {\color[RGB]{255,172,163} $\bullet$}). The remaining configurations (shown in {\color{gray!80} $\bullet$}) are not considered to be faulty.





\subsection{Manual Study}
\label{sect:manual_study}


To further understand the nature of \nfps, we explored the NVIDIA developer forums and analyzed 47 reported issues for different  ML systems (e.g., image recognition, self-driving cars, IoT, etc.). The reports and corresponding discussions highlight the challenges that  \emph{developers} and \emph{users} face when building complex systems and how they are remedied. 
Specifically, we studied composed systems deployed on NVIDIA’s hardware platforms for various applications such as image processing, voice recognition, and speech recognition deployed on heterogeneous hardware platforms (\txone, \txtwo, \xavier) allowing users to compose a custom system to meet specific functional and non-functional needs. The extensibility of composed systems exposes them to frequent misconfigurations resulting in non-functional faults. Analyzing the issues and discussions around these misconfigurations, we found the following:
%
%
\begin{enumerate*}
\item 
\textbf{There are different types of \nfps 
across all systems as a result of misconfigurations}~related to 
(i)~latency ($42\%$), (ii)~energy consumption ($36\%$), and (iii)~thermal problems ($22\%$). 
\item 
\textbf{Configuration options interact with one another in different ways across environments.}
$25\%$ of issues report multiple types of \nfps simultaneously, highlighting challenging tradeoffs users need to make between, for example, reducing latency by increasing core frequency results in higher energy consumption. 
\item 
\textbf{Non-functional faults take a long time to resolve.} Most performance issues  (\ie, 61\%) took more than 1 week (average 5 weeks, max 11 months) to resolve in comparison to other types of non-performance issues and require frequent back and forth between users and support teams, e.g.~\cite{JetsonNa94:online, PCIEx4on13:online}. Most of these issues are environment-specific and since the developers do not have access to the user's environment, they usually cannot replicate them.
\end{enumerate*}

To address these issues, developers manually inspect trace logs obtained by utilities such as
\texttt{perf}. These traces present a large amount of information pertaining to the current state of the system. The volume of information presented makes it extremely burdensome for developers to decipher the actual root-cause of non-functional faults. To overcome this challenge, we propose using causal inference accompanied by counterfactuals to reason about and address these \nfps.

%% file: 3_formulation.tex
\section{Background: Causal Inference}
\label{sect:background}

Causal inference is a systematic study to understand the interactions between variables in a system in terms of their cause-effect relationships between one another. This section provides a brief overview of how one may reason about causal relationships using the running example from example from~\fig{intro_fig}. Accordingly, let us assume that we have gathered several samples of $\text{\gpugrowth}$, $\text{\swapmem}$, and $\text{\texttt{latency}}$. If we are interested in how \texttt{latency} behaves given \gpugrowth, then, from a causal perspective, this can be formulated in three ways: observational, interventional, and counterfactual~\cite{pearl2009causality, pearl2018book}.

\subsection{Observation and Intervention}
In the \textit{observational} formulation, we measure the distribution of a target variable (\eg, latency $Y$) given that we \textit{observe} another variable \gpugrowth ($X$) takes a certain value `x' (\ie, $X=x$), denoted by $Pr(Y~|~X=x)$. This is a familiar expression that, given adequate data, modern supervised ML algorithms can estimate.

The \textit{interventional} inference tackles a harder task of estimating the {\em effects of deliberate actions}. For example, in~\fig{intro_fig}, we measure how the distribution of \texttt{Latency} ($Y$) would change if we (artificially) intervened during the data gathering process by forcing the variable \gpugrowth (X) to take a certain value `x', \ie, $X=x$, but otherwise retain the other variables (\ie, \swapmem) as is. This is done by building causal graphs and modifying them to reflect our intervention. Then, using the modified causal graph and \textit{do-causal calculus} rules~\cite{pearl2009causality} we may estimate the outcome of the artificial intervention. This process is computationally efficient since we reason over existing data instead of gathering additional measurements.

\subsection{Counterfactual Inference} 

These problems involve probabilistic answers to ``what if?'' questions.
They pertain to alternate possibilities that are counter to the current facts (hence counterfactual) and their consequences. For example, in~\fig{intro_fig}, a counterfactual question may ask:

    \begin{tabular}{!{\color{gray40}\vrule width 2pt}p{0.9\linewidth}}      
        \textit{Given that we observed a high latency, and given that \gpugrowth was set to (say) 33\%, and given everything else we know about the circumstances of the observation (\ie, available \swapmem), what is the probability of decreasing latency, had we set the \gpugrowth to 66\%?}\\
    \end{tabular}
    
\noindent In other words, we are interested in a scenario where:
\bisq
\item We \textit{hypothetically} have low latency;
\ei 
Conditioned on the following events:
\bisq
\item We \textit{actually observed} high latency;
\item \gpugrowth was initially set to $33\%$;
\item We \textit{hypothetically} set the new \gpugrowth to $66\%$; and
\item Other circumstances (available \swapmem, resource load, etc) remain the same; 
\ei
Formally we represent this, by abusing notations, as:
\begin{equation}
    \label{eq:counterfact}
    Pr(\hat{Y}=y_{low}~|~\hat{X}=0.66,~X=0.33,~Y=y_{high},~Z)
\end{equation}
Here, $Y$ stands for observed latency, $X$ stands for current \gpugrowth, and $Z$ stands for current \swapmem. The variables $\hat{X}\text{ and }\hat{Y}$ are as yet unobserved (or hypothetical), \ie, these are variables that are either being predicted for ($\hat{Y}=y_{low}$) or being \textit{hypothetically} changed to a new value ($\hat{X}=0.66$).

Questions of this nature require precise mathematical language lest they will be misleading. For example, in \eq{counterfact}, we are simultaneously conditioning on two values of \gpugrowth (\ie, $\hat{X}=0.66 \text{ and } X=0.33$). Traditional machine learning approaches cannot handle such expressions. Instead, we must resort to causal models to compute them.

Computing counterfactual questions using causal models involves three steps~\cite[Section 7]{pearl2009causality}: (i) \textit{Abduction}: where we update our model of the world using observational data; (ii) \textit{Action}: where we modify the model to reﬂect the counterfactual assumption being made; and (iii) \textit{Prediction}: where we use the obtained modified model from the previous step to compute the value of the consequence of the counterfactual. 

%% file: 4_method.tex
\section{\tool: \underline{Cau}sal \underline{Per}formance Debugging}
\label{sect:methodology}

Leveraging the Causal Analysis, described in Section~\ref{sect:background}, 
we build \tool, to debug the root causes and fix \nfps. This section presents a detailed description of \tool (outlined in ~\fig{overview}). 

\noindent
\textbf{Usage Scenario.~} In a typical scenario, \tool can be used when a developer experiences \nfps induced by misconfigurations to (a)~identify which configuration options are the \textit{root-cause} of the faults, and (b)~prescribe how to set the root-cause configuration values to fix the non-functional fault. 
To do this, the user queries \tool with their questions about the non-functional fault (\eg, ``what is the root cause of the non-functional fault?'', or ``how to improve the latency by 90\%?'', \etc).

\noindent
\textbf{Overview.}
\tool works in three phases: (I) causal structure discovery, (II) causal path extraction, and (III) debug and repair using a counterfactual query. For Phase-I, we first generate a few dozen observed data, i.e., dynamic traces for measuring the non-functional properties of the system (\eg, latency, energy consumption, etc.) under different configuration settings. Using these traces, we construct a causal graph that captures the causal relationships between various configuration options and the system's non-functional properties. For Phase-II, we use the causal graph to identify \textit{causal paths}\textemdash paths that lead from configuration options to the non- functional properties. Next, in Phase-III, given an observed \nfp, we aim to debug and repair the misconfigurations.  In particular, we generate a series of counterfactual queries, \ie, what-if questions about specific changes to the values of each configuration option. Given a \nfp as input, we check which of the counterfactual queries has the highest causal effect on remedying the \nfp, and we generate a new configuration using that query. Finally, we evaluate the new configuration to assert if the newly generated configuration mitigates the fault. If not, we add this new configuration to the observational data of Phase-I and repeat the process until the \nfp is fixed.

\input{figures/CAUPER_overview}

\input{figures/fci_samp.tex}


\subsection{Phase-I: Causal Graph Discovery}
\label{sect:structure_discovery}
In this stage, we express the relationships between configuration options (\eg, CPU freq, \etc.) and the non-functional properties (\eg, latency, \etc) using a {causal model}. A causal model is an \textit{acyclic directed mixed graph} (hereafter, ADMG), \ie, an acyclic graph consisting of directed and bidirected edges representing the causal direction and existence of latent common cause(s), respectively~\cite{richardson2002ancestral, evans2014markovian}. The nodes of the ADMG have the configuration options and the non-functional properties (\eg, latency, \etc). Additionally, we enrich the causal graph by including nodes that represent the status of internal systems events, \eg, resource pressure (as in~\fig{intro_fig}). Unlike configuration options, these system events cannot be modified. However, they can be observed and measured to explain how the causal-effect of changing configurations propagates to latency or energy consumption, \eg, in~\fig{intro_fig} resource pressure is a system event that determines how \gpugrowth affects latency.


\subsubsection*{\textbf{\em Data collection.}}~To build the causal model we gather a set of initial observational data resembling Table~\ref{tab:fci_samp_a}\footnote{For our systems, we found that 25 initial samples were adequate to build an initial version of a causal model. However, this is a parameter that may change with the system.}. It consists of configuration options set to {randomly} chosen values; various system events occurring when these configurations were used; and the measure of latency and energy consumption.



\subsubsection*{\textbf{\em Fast Causal Inference.}}~To convert observational data into a causal graph, we use a prominent structure discovery algorithm called \textit{Fast Causal Inference} (hereafter, FCI)~\cite{spirtes2000causation}. We picked FCI because it has several useful properties. First, it accommodates for the existence of unobserved confounders~\cite{spirtes2000causation,ogarrio2016hybrid, glymour2019review}, \ie, it operates even when there are unknown variables that have not been, or cannot be, measured. This is important because we do not assume absolute knowledge about configuration space, hence there could be certain configurations we could not modify or system events we have not observed. Second, it accommodates variables that belong to various data types such as nominal, ordinal, and categorical data common across the system stack.

FCI operates in three stages. First, we construct a fully connected undirected graph where each variable is connected to every other variable. Second, we use statistical independence tests to prune away edges between variables that are independent of one another. 
Finally, we orient undirected edges using prescribed edge orientation rules~\cite{spirtes2000causation,ogarrio2016hybrid, glymour2019review,colombo2012learning,colombo2014order} to produce a \textit{partial ancestral graph} (or PAG). A PAG contains the following types of (partially) directed edges: 
\bi[leftmargin=*, topsep=0pt]
\item $X$\edgeone$Y$ indicating that vertex $X$ causes $Y$. 
\item $X$\edgetwo$Y$ which indicates that there are unmeasured confounders between vertices $X$ and $Y$.
\ei
\noindent In addition, FCI produces two types of partially directed edges:
\bi[leftmargin=*, topsep=0pt]
\item $X$\edgethree$Y$ indicating that either $X$ causes $Y$, or that there are \textit{unmeasured confounders} that cause both $X$ and $Y$.
\item $X$\edgefour~$Y$ which indicates that either: (a) vertices $X$ causes $Y$, or (b) vertex $Y$ causes $X$, or (c) there are \textit{unmeasured confounders} that cause both $X$ and $Y$.
\ei

\noindent In the last two cases, the circle ($\circ$) indicates that there is an ambiguity in the edge type. In other words, given the current observational data, the circle can indicate an arrow head (\edgeone) or no arrow head (---), \ie, for $X$\edgefour$Y$, all three of $X$\edgeone$Y$, $Y$\edgeone$X$, and $X$\edgetwo$Y$ might be compatible with current data, \ie, the current data could be faithful to each of these statistically equivalent causal graphs inducing the same conditional independence relationships.


\subsubsection*{\textbf{\em Resolving partially directed edges.}}~
For subsequent analyses over the causal graph, the PAG obtained must be fully resolved (directed with no $\circ$ ended edges) in order to generate an ADMG, \ie, we must fully orient partially directed edges by replacing the circles in \edgethree and \edgefour with the correct edge direction.

Resolving these partial edges from PAG is an open problem. This is tackled in two ways: (1) Solicit expert advice to orient edges~\cite{he2008active}, but this may be too cumbersome especially when the graph is complex~\cite{he2008active}; or (2) Use alternative information-theoretic approaches to discover the correct orientations~\cite{xinpeng2020challenges, kline2015principles}. 
In this paper, we use the information-theoretic approach, enabling us to orient all edges automatically without any expert intervention. Specifically, this paper uses an entropic approach proposed in \cite{Kocaoglu2017,Kocaoglu2020} to discover the true causal direction between two variables. Entropic causal discovery is inspired by Occam’s razor, and the key intuition is that, among the possible orientations induced by partially directed edges (\ie, \edgethree and \edgefour), the most plausible orientation is that which has the lowest entropy. 

Our work extends the theoretic underpinnings of entropic causal discovery to generate a fully directed causal graph by resolving the partially directed edges produced by FCI. For each partially directed edge, we follow two steps: (1) establish if we can generate a latent variable (with low entropy) to serve as a common cause between two vertices; (2) if such a latent variable does not exist, then pick the causal direction which has the lowest entropy. 

For the first step, we assess if there could be an unmeasured confounder (say $Z$) that lies between two partially oriented nodes (say $X$ and $Y$). For this, we use the \textit{LatentSearch} algorithm proposed by Kocaoglu \etal~\cite{Kocaoglu2020}. \textit{LatentSearch} outputs a joint distribution $q(X, Y, Z)$ of the variables $X$, $Y$, and $Z$ which can be used to compute the entropy $H(Z)$ of the unmeasured confounder $Z$. Following the guidelines of Kocaoglu \etal, we set an entropy threshold $\theta_r=0.8 \times min\left\{H(X), H(Y)\right\}$. If the entropy $H(Z)$ of the unmeasured confounder falls \textit{below} this threshold, then we declare that there is a simple unmeasured confounder $Z$ (with a low enough entropy) to serve as a common cause between $X$ and $Y$ and accordingly, we replace the partial edge with a bidirected (\ie, \edgetwo) edge. 

When there is no latent variable with a sufficiently low entropy, there are two possibilities: \textit{(a)} variable $X$ causes $Y$; then, there is an arbitrary function $f(\cdot)$ such that $Y=f(X,E)$, where $E$ is an exogenous variable (independent of $X$) that accounts for noise in the system; or \textit{(b)} variable $Y$ causes $X$; then, there is an arbitrary function $g(\cdot)$ such that $X=g(Y,\tilde{E})$, where $\tilde{E}$ is an exogenous variable (independent of $Y$) that accounts for noise in the system. The distribution of $E$ and $\tilde{E}$ can be inferred from the data~\cite[see~\S3.1]{Kocaoglu2017}. With these distributions, we measure the entropies $H(E)$ and $H(\tilde{E})$. If $H(E) < H(\tilde{E})$, then, it is simpler to explain the $X$\edgeone$Y$ (\ie, the entropy is lower when $Y=f(X,E)$) and we choose $X$\edgeone$Y$ as the causal direction. Otherwise, we choose $Y$\edgeone$X$.

\subsubsection*{\textbf{Example.}}~\tab{fci_example} shows the steps involved in generating the final ADMG from~\fig{intro_fig}\hspace{-2.5pt}c. First, we build a dense graph by connecting all pairs of variables with an undirected egde (see~\fig{fci_samp_b}). Next, we use Fisher's exact test~\cite{connelly2016fisher} to evaluate the independence of all pairs of variables conditioned on all remaining variables. Pruning edges between the independent variables results in skeleton graph as shown in~\fig{fci_samp_c}. Next, we orient undirected edges using edge orientation rules~\cite{spirtes2000causation,ogarrio2016hybrid, glymour2019review,colombo2012learning} to produce a partial ancestral graph (as in~\fig{fci_samp_d}). In our example, we identify that there are two edges that are partially oriented (outlined in \red{red} in~\fig{fci_samp_d}): \textit{(i)} \swapmem \edgefour \gpugrowth; and \textit{(ii)} \texttt{resource} \texttt{pressure} \edgethree \texttt{latency}. To resolve these two edge, we use the entropic orientation strategy. The steps involved to orient these edges are illustrated in \fig{fci_samp_e}. After this step, we get the final ADMG shown in~\fig{fci_samp_f}. 

\looseness=-1

\subsection{Phase-II: Causal Path Extraction}
\label{sect:path_discovery}



In this stage, we extract paths from the causal graph (referred to as \textit{causal paths}) and rank them from highest to lowest based on their average causal effect on latency, energy consumption, and heat dissipation (our three non-functional properties). Using path extraction and ranking, we reduce the complex causal graph into a small number of useful causal paths for further analyses. The configurations in this path are more likely to be associated with the root cause of the fault.

\noindent\textbf{Extracting causal paths with backtracking.}~A causal path is a directed path originating from either the configuration options or the system event and terminating at a non-functional property (\ie, latency and/or energy consumption). To discover causal paths, we backtrack from the nodes corresponding to each non-functional property until we reach a node with no parents. If any intermediate node has more than one parent, then we create a path for each parent and continue backtracking on each parent.

%

\noindent\textbf{Ranking causal paths.~}~A complex causal graph can result in a large number of causal paths. It is not practical to reason over all possible paths as it may lead to a combinatorial explosion. Therefore, we rank the paths in descending order from ones having the highest causal effect to ones having the lowest causal effect on each non-functional property. For further analysis, we use paths with the highest causal effect.
To rank the paths, we measure the causal effect of changing the value of one node (say \gpugrowth or $X$) on its successor in the path (say \swapmem or $Z$). We express this with the \textit{do-calculus} notation: $\mathbb{E}[Z~|~\mathit{do}(X=x)]$. This notation represents the expected value of $Z$ (\swapmem) if we set the value of the node $X$ (\gpugrowth) to $x$. To compute the \textit{average causal effect} (ACE) of $X\rightarrow Z$ (\ie, \gpugrowth \edgeone \swapmem), we find the average effect over all permissible values of $X$ (\gpugrowth), \ie, 

{\footnotesize
\setlength{\abovedisplayskip}{1pt}
\setlength{\belowdisplayskip}{1pt}
\begin{multline}
\label{eq:ace}
\mathrm{ACE}\left(Z, X\right) = \frac{1}{N}\cdot \sum_{\forall a, b\in X}\mathbb{E}\left[Z~|~\mathit{do}\left(X=b\right)\right]~-~ \mathbb{E}\left[Z~|~\mathit{do}\left(X=a\right)\right]
\end{multline}
}

Here, $N$ represents the total number of values $X$ (\gpugrowth) can take. If changes in \gpugrowth result in a large change in \swapmem, then the $\mathrm{ACE}\left(Z, X\right)$ will be larger, indicating that \gpugrowth on average has a large causal effect on \swapmem. 
%
For the entire path, we extend \eq{ace} as:
\smalleq
\label{eq:path_ace}
{\footnotesize
\setlength{\abovedisplayskip}{1pt}
\setlength{\belowdisplayskip}{1pt}
\mathrm{Path}_{ACE} = \frac{1}{K} \cdot \sum \mathrm{ACE}(Z, X) \hspace{2em} \footnotesize \forall X, Z \in path 
}
\eeq
\eq{path_ace} represents the average causal effect of the causal path. The configuration options the lie in paths with larger $P_{ACE}$ tend to have a greater causal effect on the corresponding non-functional properties in those paths. We select the top $K$ paths with the largest $\mathrm{P}_{ACE}$ values, for each non-functional property. In this paper, we use K=3, however, this may be modified in our replication package. 


\subsection{Phase-III: Debugging non-functional faults}
\label{sect:root_cause}

In this stage, we use the top $K$ paths to (a)~identify the root cause of non-functional faults; and (b)~prescribe ways to fix the non-functional faults.
When experiencing non-functional faults, a developer may ask specific queries to \tool and expect an actionable response. For this, we translate the developer's queries into formal probabilistic expressions that can be answered using causal paths. We use counterfactual reasoning to generate these probabilistic expressions. 
To understand query translation, we use the example causal graph of~\fig{fci_samp_f} where a developer observes high latency, \ie, a latency fault, and has the following questions:  

\noindent\textbf{\faQuestionCircle~\textbf{``What is the root cause of my latency fault?''}} To identify the root cause of a non-functional fault we must identify which configuration options have the most causal effect on the performance objective. 
For this, we use the steps outlined in~\tion{path_discovery} to extract the paths from the causal graph and rank the paths based on their average causal effect (\ie, $\mathrm{Path}_{ACE}$ from \eq{path_ace}) on latency. We return the configurations that lie on the top $K$ paths. 
For example, in~\fig{fci_samp_f} we may return the path (say) \gpugrowth~\edgeone \swapmem \edgeone Latency and the configuration options {\gpugrowth \textit{and} \swapmem} both being probable root causes.

\noindent\textbf{\faQuestionCircle~\textbf{``How to improve my latency?''}} To answer this query, we first find the root cause as described above. Next, we discover what values each of the configuration options must take in order that the new latency is better (low latency) than the fault (high latency). For example, we consider the causal path \gpugrowth~\edgetwo \swapmem \edgeone Latency, we identify the permitted values for the configuration options {\gpugrowth and \swapmem} that can result in a low latency ($Y^{\mathit{\textsc{low}}}$) that is better than the fault ($Y^{\mathit{\textsc{high}}}$).
For this, we formulate the following counterfactual expression: 
\smalleq
\label{eq:cfact_bare}
\footnotesize
\mathrm{Pr}(Y_{repair}^{\textsc{low}}|\neg repair,Y_{\neg repair}^{\textsc{high}})
\eeq
\eq{cfact_bare} measures the probability of ``fixing'' the latency fault with a ``repair'' {\footnotesize $(Y_{repair}^{\textsc{low}})$} given that with no repair {we observed the fault} {\footnotesize $(Y_{\neg repair}^{\text{\textsc{high}}})$}.   
In our example, the repairs would resemble \gpugrowth=$0.66$ or \swapmem=$4Gb$, \etc. We generate a \textit{repair set} ($\mathcal{R}$) where the configurations \gpugrowth and \swapmem are set to all permissible values, \ie,
{
\small
\begin{multline}\label{eq:repairs}
    \setlength{\abovedisplayskip}{5pt}
    \setlength{\belowdisplayskip}{5pt}
    \mathcal{R}\equiv~\bigcup~\left\{\text{\gpugrowth} = {x},~\text{\swapmem} = {z},... \right\}\\\forall {x} \in \text{\gpugrowth},~{z} \in \text{\swapmem},~\ldots
\end{multline}}
Next, we compute the \textit{individual treatment effect} (ITE) on the latency ($Y$) for each repair in the repair set $\mathcal{R}$. In our case, for each repair $\mathit{r}~\in~\mathcal{R}$, ITE is given by:
\begin{equation}
    \label{eq:ite}
    \footnotesize
    \mathrm{ITE}(\mathit{r})=\mathrm{Pr}(Y_r^{\textsc{low}}~|~\neg r,~Y_{\neg r}^{\textsc{high}}) - \mathrm{Pr}(Y_r^{\textsc{high}}~|~\neg r,~Y_{\neg r}^{\textsc{high}})\hspace{1em}
\end{equation}
ITE measures the difference between the probability that the latency \textit{is low} after a repair $r$ and the probability that the latency is \textit{still high} after a repair $r$. If this difference is positive, then the repair has a higher chance of fixing the fault. In contrast, if the difference is negative then that repair will likely worsen the latency. To find the most useful repair ($\mathcal{R}_{\mathit{best}}$), we find a repair with the largest (positive) ITE, \ie, 
$$\mathcal{R}_{\mathit{best}} = \argmax_{\forall r~\in~\mathcal{R}}[\mathrm{ITE}(\mathit{r})]$$

This provides the developer with a possible repair for the configuration options that can fix the latency fault.

\subsubsection*{Remarks.}~The ITE computation of \eq{ite} occurs \textit{only} on the observational data. Therefore we may generate any number of repairs and reason about them without having to deploy those interventions and measuring their performance in the real-world. This offers significant monetary and runtime benefits.

\subsection{Incremental Learning}
\label{sect:incremental_learning}
Using the output of \textbf{Phase-III}, the system is reconfigured with the new configuration. If the new configuration addresses the \nfp, we return the recommended repairs to the developer. 
Since the causal model uses limited observational data, there may be a discrepancy between the actual performance of the system after the repair and the value of the estimation derived from the current version of the causal graph. The more accurate the causal graph, the more accurate the proposed intervention will be~\cite{spirtes2000causation,ogarrio2016hybrid, glymour2019review,colombo2012learning,colombo2014order}.
Therefore, in case our repairs do not fix the faults, we update the observational data with this new configuration and repeat the process. Over time, the estimations of causal effects will become more accurate. We terminate the incremental learning once we achieve the desired performance.

%% file: figures/CAUPER_overview.tex
\begin{figure}[t!]
    \centering
    \includegraphics*[width=\linewidth]{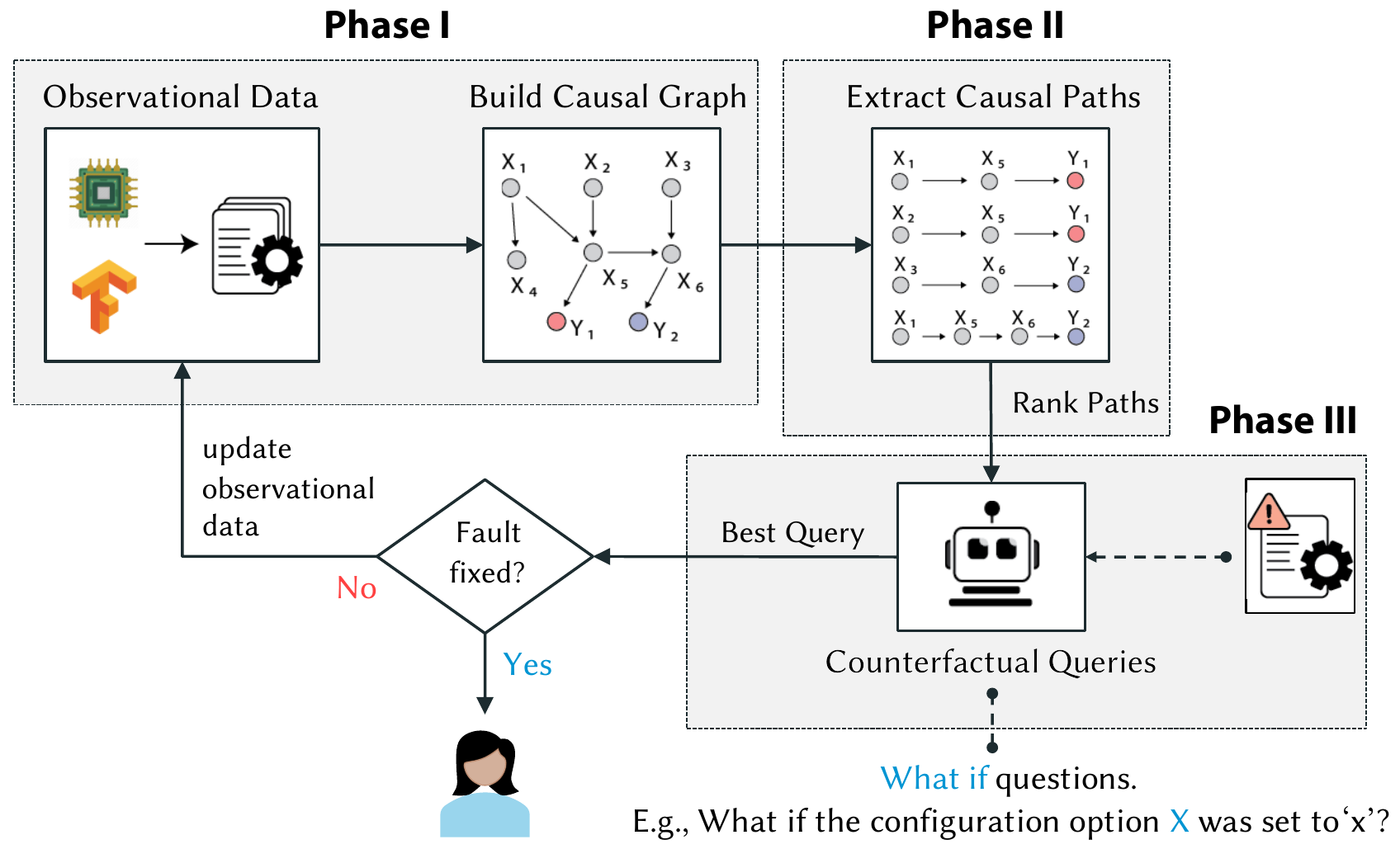}
    \caption{\small {Overview of \tool}}
    \label{fig:overview}
\end{figure}

%% file: figures/fci_samp.tex
\begin{figure*}[tbp!]
    \subfloat[Sample Observational Data]{
        \resizebox{0.33\linewidth}{!}{\adjustbox{valign=b}{%
        \begin{tabular}{@{}c|cc@{}cc|r|}
            \clineB{2-6}{2}
        \multicolumn{1}{c|}{} & \multicolumn{2}{c}{\textit{(Configurable Options)}} & &  \multicolumn{1}{c|}{\textit{(Sys. Events)}} & Latency \bigstrut[t]\\
        \multicolumn{1}{c|}{} & GPU Growth & Swap Mem & &Resource Use &  \etc. \\ \hlineB{2}
        $c_1$ & $0.25$ & $2~Gb$ & & $10\%$ & \cellcolor{gray05}1 sec, \dots \\
        $c_2$ & $0.5$ & $1~Gb$ & & $20\%$ & \cellcolor{gray05}2 sec, \dots \\
        $c_3$ & $0.66$ & $1.5~Gb$ & & $40\%$ & \cellcolor{gray05}1.5 sec, \dots \\
        $c_4$ & $0.75$ & $3~Gb$ & & $60\%$ & \cellcolor{gray05}3 sec, \dots \\
        $\vdots$ & $\vdots$ & $\vdots$ & \multicolumn{1}{@{}c@{}}{$\ddots$} & $\vdots$ & \multicolumn{1}{c|}{\cellcolor{gray05}$\vdots$} \\
        $c_n$ & $1.0$ & $4~Gb$ & & $40\%$ & \cellcolor{gray05}0.1 sec, \dots \\ \hlineB{2}
        \end{tabular}}}
        \label{tab:fci_samp_a}}
 \qquad\subfloat[Dense Graph]{\includegraphics*[width=0.155\linewidth, valign=b]{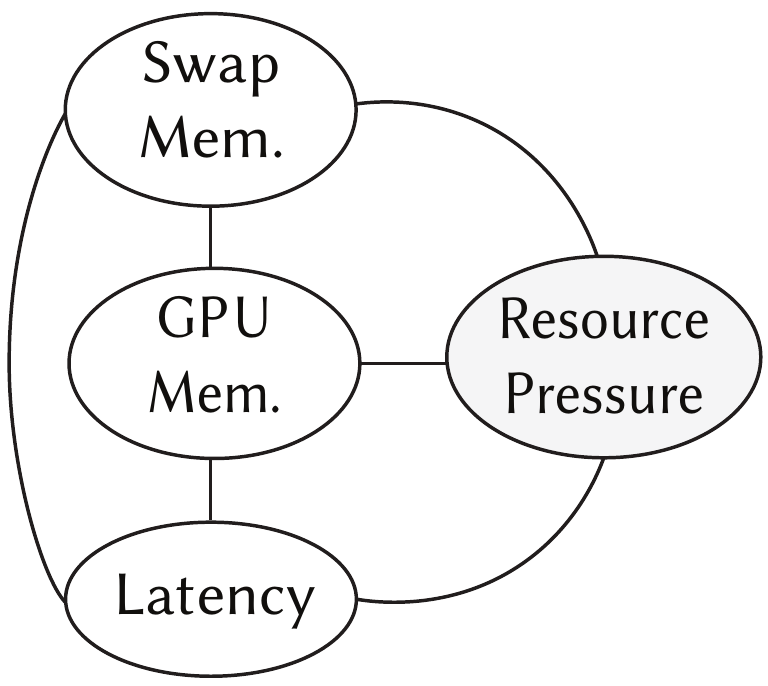}\label{fig:fci_samp_b}} 
 \qquad
 \subfloat[FCI Skeleton]{\includegraphics*[width=0.145\linewidth, valign=b]{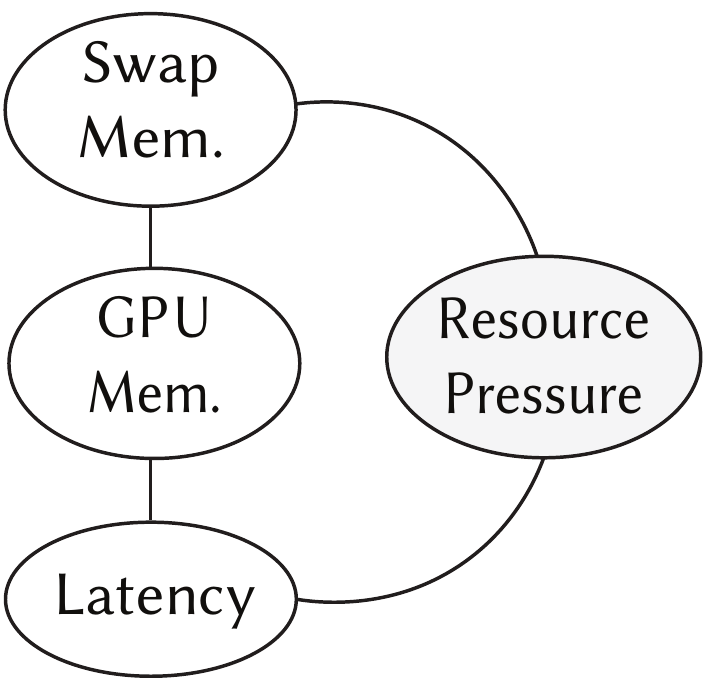}\label{fig:fci_samp_c}}
 \qquad
 \subfloat[Partial Ancestral Graph (PAG) from FCI]{\includegraphics*[width=0.145\linewidth]{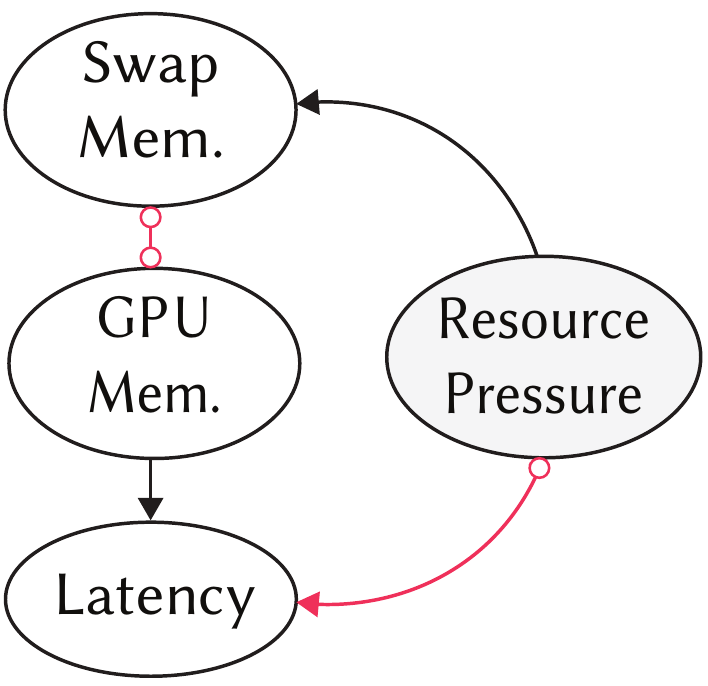}\label{fig:fci_samp_d}}
 \vspace{-.6cm}
 \subfloat[Orienting partially directed edges from \protect\fig{fci_samp_d} using entropy]{\includegraphics*[width=0.8\linewidth, valign=b]{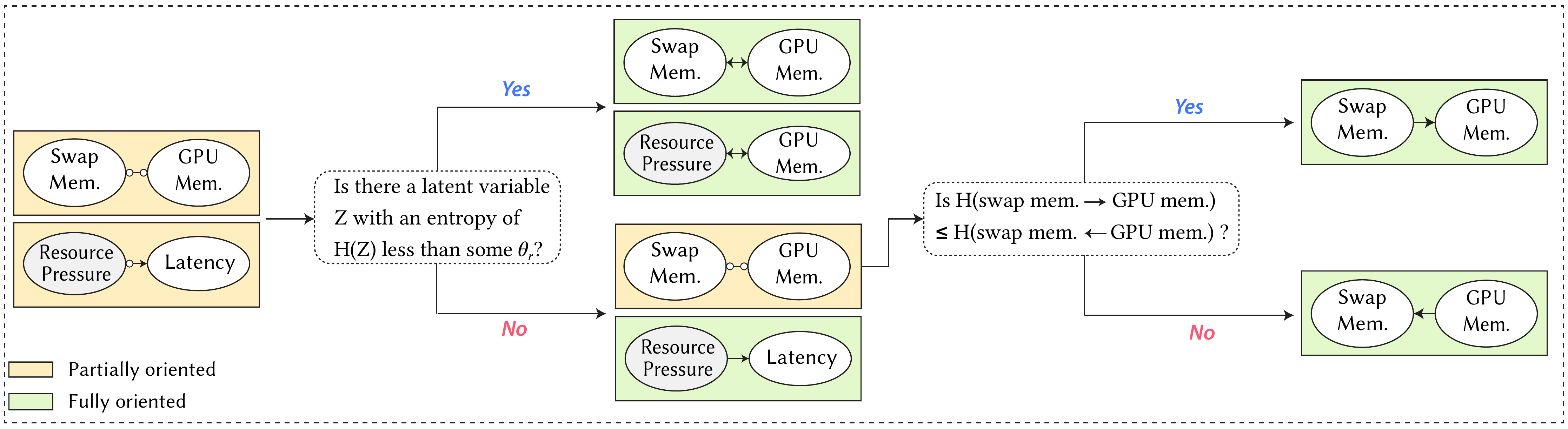}\label{fig:fci_samp_e}}
\qquad
 \subfloat[Final ADMG]{\includegraphics*[width=0.145\linewidth, valign=b]{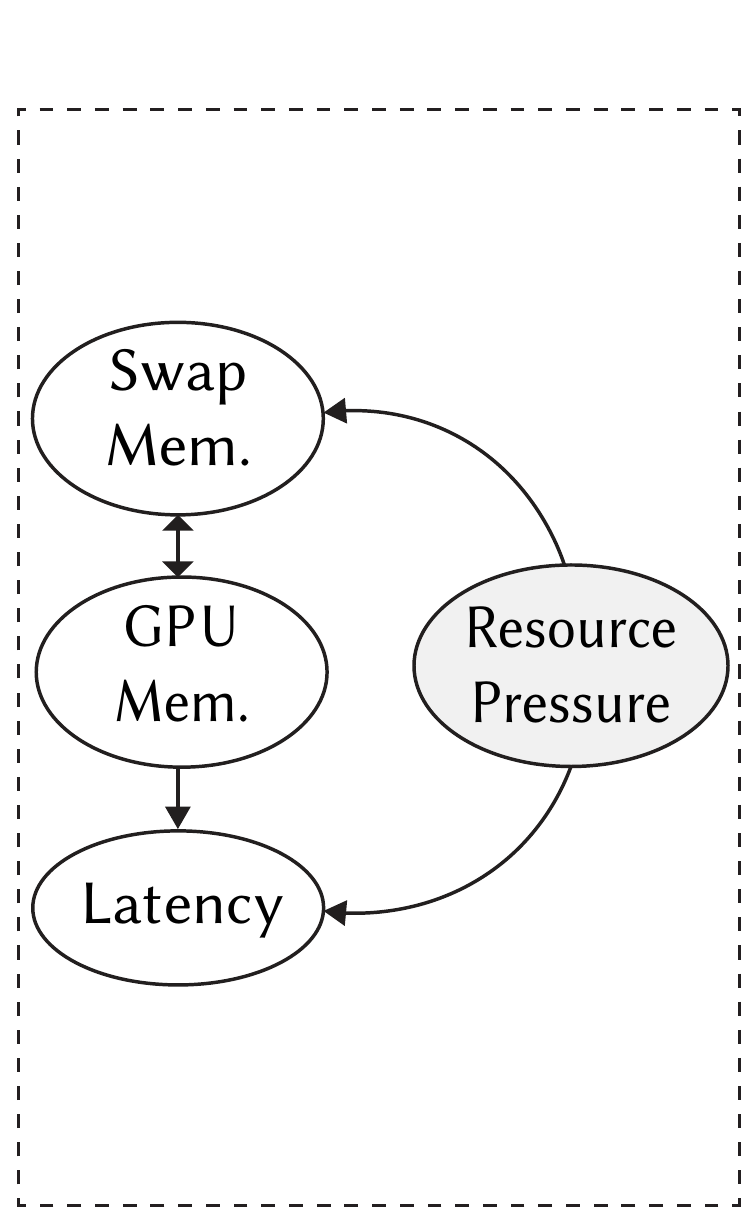}\label{fig:fci_samp_f}}
\caption{\small {From observational data to a fully connected graph, a skeleton graph, and finally to partial ancestral graph (PAG).}}

\label{tab:fci_example}\end{figure*}

%% file: 11_case_study.tex
\section{Case Study: Latency Fault in \txtwo}

\begin{figure}[t!]
    \setlength{\belowcaptionskip}{-1em}
    \centering
    \resizebox{0.965\linewidth}{!}{
        \begin{tabular}{V{2.5}p{\linewidth}V{2.5}}
            \hlineB{2}
            \small
            \textbf{Problem~\cite{code_transplant:online}:}~For a real-time scene detection task, \txtwo (faster platform) only processed 4 frames/sec whereas \txone (slower platform) processed 17 frames/sec, \ie, the latency is $4\times$ worse on \txtwo.
            \\
            \small\textbf{Observed Latency (frames/sec):} 4 FPS\\
            \small\textbf{Expected Latency (frames/sec):} 22-24 FPS \textit{(30-40\% better)}\\\hlineB{2}
        \end{tabular}
    }\vspace{0.1em}
    \resizebox{0.95\linewidth}{!}{%
    \begin{tabular}{lV{2.5}ccccV{2.5}r}
    \clineB{2-6}{2.5}
     \multicolumn{1}{lV{2.5}}{\textbf{Configuration Options}}& \rotatebox{90}{\tool~} & \rotatebox{90}{SMAC} & \rotatebox{90}{\bugdoc~} & \rotatebox{90}{Forum} & \multicolumn{1}{cV{2.5}}{\rotatebox{90}{ACE$^\dagger$}} \bigstrut[t]\\ \clineB{2-6}{2.5}
     
    \multicolumn{1}{l}{} & \multicolumn{1}{l}{} & \multicolumn{1}{l}{} & \multicolumn{1}{l}{} & \multicolumn{1}{l}{} & \multicolumn{1}{l}{} \\[-0.9em] \hlineB{2.5}
     
    \multicolumn{1}{V{2.5}lV{2.5}}{CPU Cores} & \cellcolor{blue!10}\color{gray50}{\faCheck} & \cellcolor{blue!10}\color{gray50}{\faCheck} & \cellcolor{blue!10}\color{gray50}{\faCheck} & \cellcolor{blue!10}\color{gray50}{\faCheck} & \multicolumn{1}{cV{2.5}}{3\%} \bigstrut[t]\\

    \multicolumn{1}{V{2.5}lV{2.5}}{CPU Frequency} & \cellcolor{blue!10}\color{gray50}{\faCheck} & \cellcolor{blue!10}\color{gray50}{\faCheck} & \cellcolor{blue!10}\color{gray50}{\faCheck} & \cellcolor{blue!10}\color{gray50}{\faCheck} & \multicolumn{1}{cV{2.5}}{6\%} \\

    \multicolumn{1}{V{2.5}lV{2.5}}{EMC Frequency} & \cellcolor{blue!10}\color{gray50}{\faCheck} & \cellcolor{blue!10}\color{gray50}{\faCheck} & \cellcolor{blue!10}\color{gray50}{\faCheck} & \cellcolor{blue!10}\color{gray50}{\faCheck} & \multicolumn{1}{cV{2.5}}{13\%} \\

    \multicolumn{1}{V{2.5}lV{2.5}}{GPU Frequency} & \cellcolor{blue!10}\color{gray50}{\faCheck} & \cellcolor{blue!10}\color{gray50}{\faCheck} & \cellcolor{blue!10}\color{gray50}{\faCheck} & \cellcolor{blue!10}\color{gray50}{\faCheck} & \multicolumn{1}{cV{2.5}}{22\%} \\

    \multicolumn{1}{V{2.5}lV{2.5}}{Scheduler Policy} & $\cdot$ & \cellcolor{orange!12}\color{gray50}{\faCheck} & \cellcolor{orange!12}\color{gray50}{\faCheck} & $\cdot$ & \multicolumn{1}{cV{2.5}}{.} \\

    \multicolumn{1}{V{2.5}lV{2.5}}{Sched rt runtime} & $\cdot$ & $\cdot$ & \cellcolor{orange!12}\color{gray50}{\faCheck} & $\cdot$ & \multicolumn{1}{cV{2.5}}{.} \\

    \multicolumn{1}{V{2.5}lV{2.5}}{Sched child runs} & $\cdot$ & $\cdot$ & \cellcolor{orange!12}\color{gray50}{\faCheck} & $\cdot$ & \multicolumn{1}{cV{2.5}}{.} \\

    \multicolumn{1}{V{2.5}lV{2.5}}{Dirty bg. Ratio} & $\cdot$ & $\cdot$ & $\cdot$ & $\cdot$ & \multicolumn{1}{cV{2.5}}{.} \\

    \multicolumn{1}{V{2.5}lV{2.5}}{Dirty Ratio} & $\cdot$ & $\cdot$ & \cellcolor{orange!12}\color{gray50}{\faCheck} & $\cdot$ & \multicolumn{1}{cV{2.5}}{.} \\

    \multicolumn{1}{V{2.5}lV{2.5}}{Drop Caches} & $\cdot$ & \cellcolor{orange!12}\color{gray50}{\faCheck} & \cellcolor{orange!12}\color{gray50}{\faCheck} & $\cdot$ & \multicolumn{1}{cV{2.5}}{.} \\

    \multicolumn{1}{V{2.5}lV{2.5}}{\texttt{CUDA\_STATIC}} & \cellcolor{blue!10}\color{gray50}{\faCheck} & \cellcolor{blue!10}\color{gray50}{\faCheck} & \cellcolor{blue!10}\color{gray50}{\faCheck} & \cellcolor{blue!10}\color{gray50}{\faCheck} & \multicolumn{1}{cV{2.5}}{55\%} \\

    \multicolumn{1}{V{2.5}lV{2.5}}{Cache Pressure} & $\cdot$ & $\cdot$ & $\cdot$ & $\cdot$ & \multicolumn{1}{cV{2.5}}{.} \\

    \multicolumn{1}{V{2.5}lV{2.5}}{Swappiness} & $\cdot$ & \cellcolor{orange!12}\color{gray50}{\faCheck} & \cellcolor{orange!12}\color{gray50}{\faCheck} & $\cdot$ & \multicolumn{1}{cV{2.5}}{1\%}\\ \hlineB{2.5}

    \multicolumn{1}{l}{} & \multicolumn{1}{l}{} & \multicolumn{1}{l}{} & \multicolumn{1}{l}{} & \multicolumn{1}{l}{} & \multicolumn{1}{l}{} \\[-0.95em] \clineB{1-5}{2.5}
     
    \multicolumn{1}{V{2.5}lV{2.5}}{Latency (\txtwo frames/sec)} & \textbf{26} & 24 & 20 & \multicolumn{1}{lV{2.5}}{23} &  \bigstrut[t]\\
    \multicolumn{1}{V{2.5}lV{2.5}}{Latency Gain (over \txone)} & \textbf{53\% } & 42\% & 21\% & \multicolumn{1}{lV{2.5}}{39\%} &  \\
    \multicolumn{1}{V{2.5}lV{2.5}}{Latency Gain (over default)} & \textbf{6.5$\times$} & 6$\times$ & 5$\times$ & \multicolumn{1}{lV{2.5}}{5.75$\times$} &  \\
    \multicolumn{1}{V{2.5}lV{2.5}}{Resolution time} & \textbf{24 mins} & 4 hrs & 3.5 hrs & \multicolumn{1}{lV{2.5}}{2 days} &  \\ \clineB{1-5}{2.5}
\end{tabular}}%
\vspace{1em}
\includegraphics[width=0.95\linewidth]{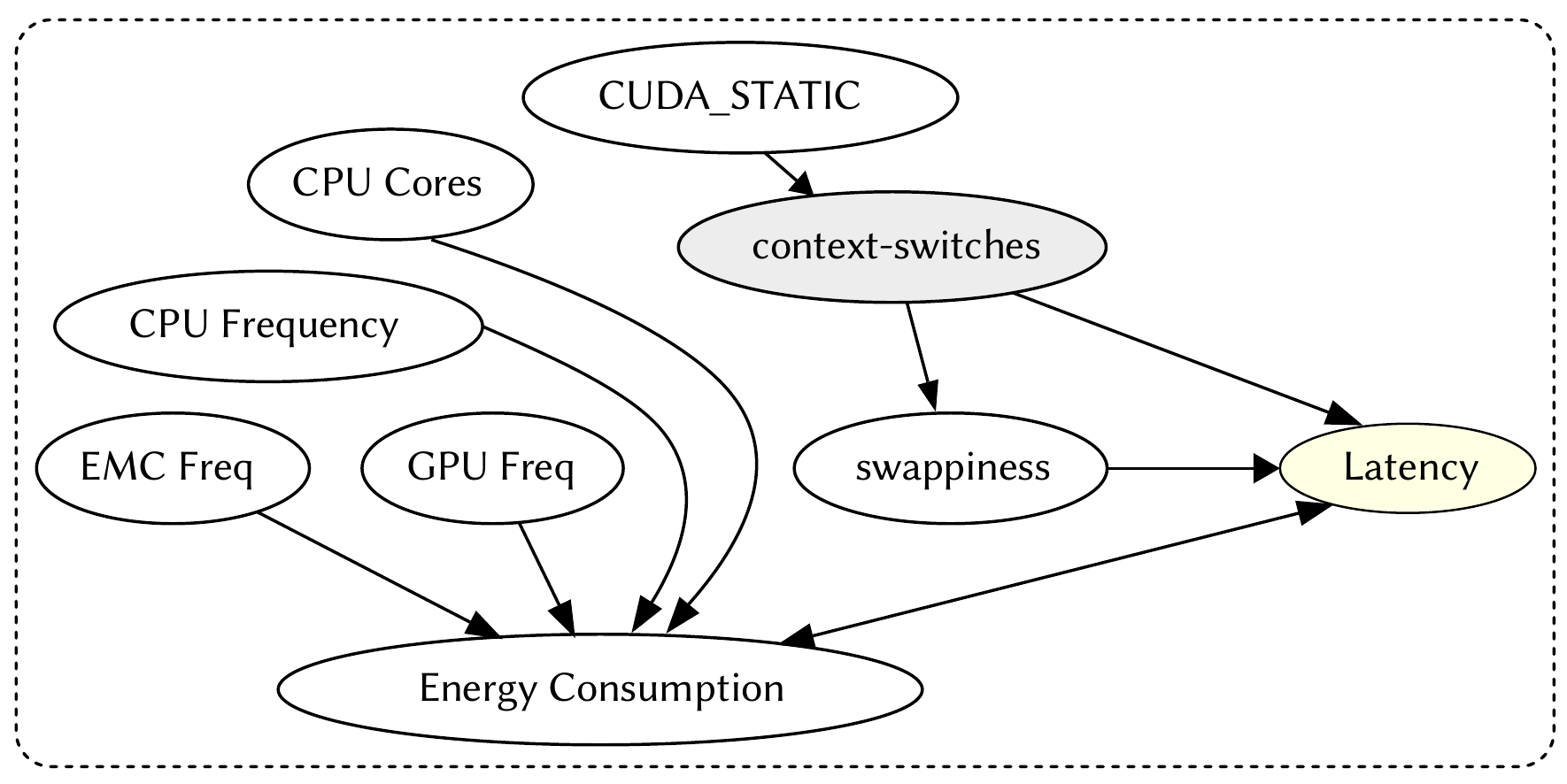}

\caption{\small Using \tool on the real-world example from~\tion{motivating_examples}. \tool is better and faster than other methods.}
\label{fig:real_example}
\end{figure}

This section revisits the real-world latency fault previously discussed in \tion{motivating_examples}. 
For this study, we reproduce the developers' setup to assess how effectively \tool can diagnose the root-cause of the misconfigurations and fix them. For comparison, we use SMAC (an optimization approach) and \bugdoc (an ML-based diagnosis tool). We evaluate the methods against the recommendations by the domain experts on the forum.
\subsubsection*{Findings.~}~%
\fig{real_example} illustrates our findings. We find that:
\bisq
\item \tool could diagnose the root-cause of the misconfiguration and recommends a fix within 24 minutes. Using the recommended configuration fixes from \tool, we achieved a frame rate of 26 FPS ($53\%$ better than \txone and $6.5\times$ better than the fault). This exceeds the developers' initial expectation of $30-40\%$ improvement. 
\item Using configuration optimization (SMAC), we auto-tune the system to find a near-optimal configuration. This (near-optimal) configuration had a latency of 24 FPS (which was $42\%$ better than \txone and $6\times$ better than the fault). While SMAC also meets the developer's expectations, it performed slightly worse than \tool, and took 4 hours ($6\times$ longer than \tool). Further, the near-optimal configuration found by SMAC, change several unrelated configurations which were not recommended by the experts (\colorbox{orange!12}{\color{gray50}{\faCheck}} in~\fig{real_example}). 
\item \bugdoc (ML-based approach) has the least improvement compared to other approaches ($21\%$ improvement over \txone) while taking 3.5 hours (mostly spent on collecting training samples to train internal the decision tree.). \bugdoc failed to meet the developer's expectation. \bugdoc also changed several unrelated configurations (depicted by \colorbox{orange!12}{\color{gray50}{\faCheck}}) not endorsed by the domain experts.   
\ei

\subsubsection*{Why \tool works better (and faster)?~}~%
\tool discovers the misconfigurations by constructing a causal model (a simplified version of this is shown in \fig{real_example}). This causal model rules out irrelevant configuration options and focuses on the configurations that have the highest (direct or indirect) causal effect on latency, \eg, we found the root-cause \texttt{CUDA} \texttt{STATIC} in the causal graph which indirectly affects latency via context-switches (an intermediate system event); this is similar to other relevant configurations that indirectly affected latency (via energy consumption). Using counterfactual queries, \tool can reason about changes to configurations with the highest average causal effect (last column in~\fig{real_example}). The counterfactual reasoning occurs no additional measurements, significantly speeding up inference. 

Together, the causal model and the counterfactual reasoning enable \tool to pinpoint the configuration options that were misconfigured and recommend a fix for them in a timely manner. As shown in \fig{real_example}, \tool accurately finds all the configuration options recommended by the forum (depicted by \colorbox{blue!10}{\color{gray50}{\faCheck}} in~\fig{real_example}). Further, \tool recommends fixes to these options that result in $24\%$ better latency than the recommendation by domain experts in the forum. More importantly, \tool takes only 24 minutes (vs. 2 days of forum discussion) without modifying unrelated configurations. 
%


%% file: 5_evaluation.tex
\section{Experimental Setup}
\label{sect:evaluation}

\subsection{Study subjects}
\label{sect:study_subjects}
\noindent\textbf{Hardware Systems.}~This study uses three NVIDIA Jetson Platforms: \txone, \txtwo, and \xavier. Each platform has different hardware specifications, \eg, different CPU micro-architectures (ARM Caramel, Cortex), GPU micro-architectures (Maxwell, Pascal, Volta), energy requirements (5W--30W). 

\noindent\textbf{Software systems.}~
We deploy five software systems on each NVIDIA Jetson Platform:
(1)~CNN based Image recognition with Xception to classify 5000 images from the CIFAR10 dataset~\cite{chollet2017xception};
(2)~BERT (a transformer-based model) to perform sentiment analysis on 10000 reviews from the IMDb dataset~\cite{devlin2018bert};
(3)~DeepSpeech an RNN based voice recognition on 5sec long audio files~\cite{hannun2014deep};
(4)~SQLite, a database management system, to perform read, write, and insert operations; and
(5)~x264 video encoder to encode a video file of size 11MB with resolution 1920 x 1080.

\noindent\textbf{Configuration Options.}~This work uses 28 configuration options including 10 software configurations, 8 OS/Kernel configurations, and 10 hardware configurations (c.f. \fig{real_example}). We also record the status of 19 non-intervenable system events. We choose these configurations and system events based on NVIDIA's configuration guides/tutorials and other related work~\cite{Halawa2017}. 

\subsection{Benchmark Dataset}
\label{sect:jetson_faults}



We curate a non-functional faults dataset, called the {\sc Jetson Faults} dataset, for each of the software and hardware system used in our study. This dataset and the data collection scripts are available in our \href{https://bit.ly/30Q4PSC}{replication package}. 

\noindent\textbf{Data Collection.} 
We exhaustively set each configuration option to all permitted values. Note, for continuous configuration options we choose 10 equally spaced values between the minimum and maximum permissible values, \eg, for \gpugrowth, we vary the value between 0\% and 100\% in steps of 10\%. Next, we measure the latency, energy consumption, and heat dissipation for every configuration. We repeat each measurement $5$ times and record the average to handle system noise and other variabilities~\cite{iqbal2019transfer}.

\noindent\textbf{Labeling misconfigurations.}~ By definition, non-functional faults have latency, energy consumption, and heat dissipation that take tail values~\cite{gunawi2018fail,kleppmann2017designing}, \ie, they are worse than the $99^\text{th}$ percentile. We filter our data set to find the configurations that result in tail values for latency and/or energy consumption, and label these configurations as `\textit{faulty}'. 
%


\noindent\textbf{Ground Truth.}~We create a \textit{ground-truth data} by inspecting each configuration labeled faulty and identifying their root-causes manually. We also manually reconfigure each faulty configuration to achieve the best possible latency, energy consumption, and/or heat dissipation thereby dispensing with the non-functional fault
. For each reconfiguration, we record the changed configuration options and the values; these are used as a reference for evaluation.

\subsection{Baselines}
\label{sect:baselines}
\tool is compared against four alternative state of-the-art ML-based methods for fault diagnostic and mitigation: 
\begin{enumerate*}
\item \textsc{\bfseries CBI}~\cite{song2014statistical}: A correlation based feature selection algorithm for identifying and fixing fault inducing configurations; 
\item \textsc{\bfseries Delta Debugging}~\cite{artho2011iterative}: A debugging technique that minimizes the difference between a pair of configurations, where one configuration causes a fault while the other does not;
\item \textsc{\bfseries  EnCore}~\cite{zhang2014encore}: A rule based technique that learns correlational information about misconfigurations from a given set of sample configurations.
\item \textsc{ \bfseries BugDoc}~\cite{lourencco2020bugdoc}: A Debugging Decision Tree based approach to automatically infer the root causes and derive succinct explanations of failures.
\end{enumerate*}

We also compare \tool with the near-optimal configurations from two state-of-the-art configuration optimization methods: (1) \textsc{\bfseries  SMAC} \cite{hutter2011sequential}: A sequential model based optimization technique to auto-tune software systems; (2) \textsc{\bfseries  PESMO} \cite{hernandez2016predictive}: A multi-objective bayesian optimization to find Pareto-optimal configurations.

\subsection{Evaluation Metrics}\label{sect:metrics}
\noindent\textbf{Relevance scores.~}We evaluate the predicted root-causes in terms of (a) the true positive rate (recall), (b) the true discovery rate (precision), and (c) accuracy (jaccard similarity). We prefer high accuracy, precision, and recall. To compute these metrics, we compare the set of configuration options identified by \tool to be the root cause with the true root-cause from the ground truth (\tion{jetson_faults}). 

\noindent\textbf{Repair quality.~} To assess the quality of fixes, we measure the percentage improvement (gain \%) after applying the recommended repairs using $\Delta_{gain}$ defined as:
$\Delta_{gain}=\frac{\text{NFP}_\textsc{fault}-\text{NFP}_\textsc{nofault}}{\text{NFP}_\textsc{fault}}\times 100$.
%
%
Here $\text{NFP}_{\textsc{fault}}$ is the value of the faulty non-functional property (latency,\etc.) and $\text{NFP}_{\textsc{no}~\textsc{fault}}$ is the the value of the faulty non-functional property after applying the repairs recommended by \tool. The larger the $\Delta_{gain}$, the better the recommended fix.  


%% file: 6_results.tex
\section{Experimental Results}

\subsection*{RQ1. \tool vs. Model-Based Diagnostics}
\subsubsection*{\textbf{Motivation.}} Machine learning approaches are commonly used in model-based fault diagnostics~\cite{artho2011iterative, song2014statistical, lourencco2020bugdoc, zhang2014encore}. These models are trained to learn the correlation between the configuration options and the non-functional properties (e.g, latency, \etc) which are then extrapolated to diagnose and fix faults. This research question compares \tool and four state-of-the-art ML-based methods (described in~\tion{baselines}) for single objective fault diagnostics and multi-objective fault diagnostics.


%
%
%
%

\noindent\textbf{Note.~}~%
~All approaches require some initial observational data to operate. For \tool we begin with 25 initial samples to let \tool incrementally generate, evaluate, and update the causal model with candidate repairs. Other methods require a larger pool of training data. However, collecting the training data is expensive and time-consuming. Therefore, we use a budget of 3 hours to generate random configuration samples to train the other methods.

%
\subsubsection*{\textbf{Approach.}}~We assess the effectiveness of diagnostics for two types of non-functional faults: (a)~``single-objective'' faults that occur either in latency or in energy consumption, and (b)~``multi-objective'' faults where misconfigurations affect multiple non-functional properties simultaneously, \ie, in latency \textit{and} energy consumption.
For brevity, we evaluate latency faults in \txtwo and energy consumption faults in \xavier. Our findings generalize over other hardware.

\input{tables/rq1.tex}

\subsubsection*{\textbf{Approach.}}~\Cref{tab:rq1_1,tab:rq2} compare the effectiveness of \tool over other ML-based fault diagnosis approaches. We observe: 

\noindent\textbf{$\bullet~$~\textit{Accuracy, precision, and recall:}~} \tool~{\em significantly outperforms ML-based methods in all cases}. For example, in image recognition with Xception on TX2, \tool achieves 13\% more accuracy, 19\% larger Precision compared to \bugdoc and 14\% more recall compared to \encore. We observe similar trends in energy faults and multi-objective faults, \ie, \tool outperforms other methods. 


\noindent\textbf{$\bullet~$~\textit{Gain.}~}~\tool~{\em can recommend repairs for faults that significantly improve latency and energy usage}. Applying the changes to the configurations recommended by \tool increases the performance drastically. We observed latency gains as high as $83\%$ ($23\%$ more than \bugdoc) on \txtwo and energy gain of $83\%$ ($32\%$ more than \bugdoc) on \xavier for image recognition.

\noindent\textbf{$\bullet~$~\textit{Wallclock time.}~}~\tool~{\em can resolve misconfiguration faults significantly faster than ML-based approaches}.~In~\Cref{tab:rq1_1,tab:rq2}, the last two columns indicate the time taken (in hours) by each approach to diagnosing the root cause. For all methods, we set a maximum budget of 4 hours. We find that, while other approaches use the entire budget to diagnose and resolve the faults, \tool can do so significantly faster, \eg, \tool is $13\times$ faster in diagnosing and resolving faults in energy usage for x264 deployed on \xavier and $10\times$ faster for latency faults for NLP task on \txtwo. 

\subsubsection*{\textbf{Discussion.}}~Based on the above, we answer the following additional questions:

\noindent\textit{1. Why is \tool better than model-based approaches?}
Model-based methods rely only on correlation, and this can be misleading since they cannot incorporate the intrinsic complex causal structure of the underlying configuration space. In contrast, \tool relies on causal inference (instead of correlation) to model the configuration space thereby overcoming the limitations with current ML-models. 

\noindent\textit{2. Why is \tool faster than model-based approaches?}
ML-based methods require a large number of initial observational data for training. They spend most of their allocated 4-hour budget on gathering these training samples. In contrast, \tool starts with only 25 samples and uses incremental learning (\tion{incremental_learning}) to judiciously update the casual graph with new configurations until a repair has been found. This drastically reduces the inference time.

\subsection*{RQ2: \tool vs. Search-Based Optimization}
\label{sect:rq3}



\subsubsection*{\textbf{Motivation.}} Search based methods use techniques such as Bayesian optimization~\cite{snoek2012practical, shahriari2015taking} to explore the relationships between configuration options and corresponding non-functional properties. 
Albeit recent approaches seem promising~\cite{hsu2018arrow, dalibard2017boat, miyazaki2018bayesian}, these methods are not viable for diagnostics since our objective is not to find an optimal configuration, rather it is to fix an already encountered fault. In this RQ, we explore the challenges of using optimization for fault diagnostics in terms of their efficacy (RQ2-A), their analysis time (RQ2-B), and their sensitivity to external changes (RQ2-C), 
by comparing \tool with two popular optimization approaches: (a) SMAC and (b) PESMO (see~\tion{baselines}).



\begin{figure}[t]
    \setlength{\belowcaptionskip}{-1.5em}
    \subfloat[Single Objective]{
        \includegraphics[width=0.27\linewidth]{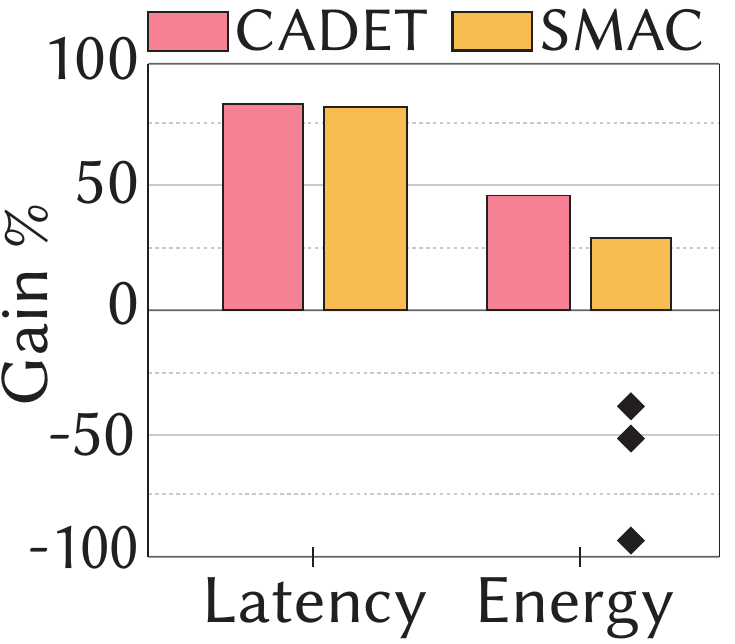}
        \label{fig:rq3_1_a}
    }
    \subfloat[Multi Objective]{
        \includegraphics[width=0.275\linewidth]{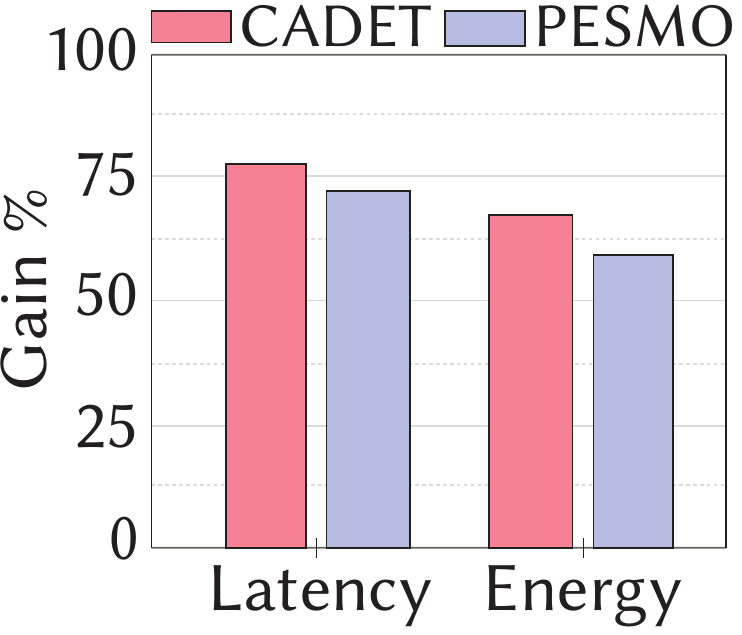}
        \label{fig:rq3_1_b}
    }
    \subfloat[Wallclock time.]{
        \includegraphics[width=0.38\linewidth]{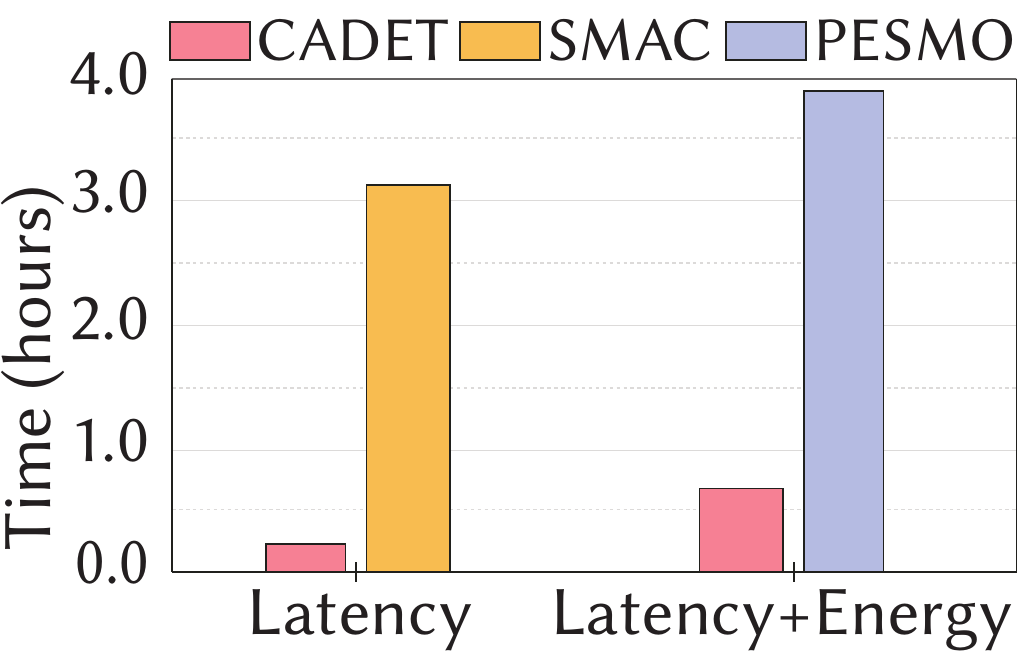}
        \label{fig:rq3_1_c}
    }
    \\
    \subfloat[Sensitivity.]{
        \includegraphics[width=0.4\linewidth]{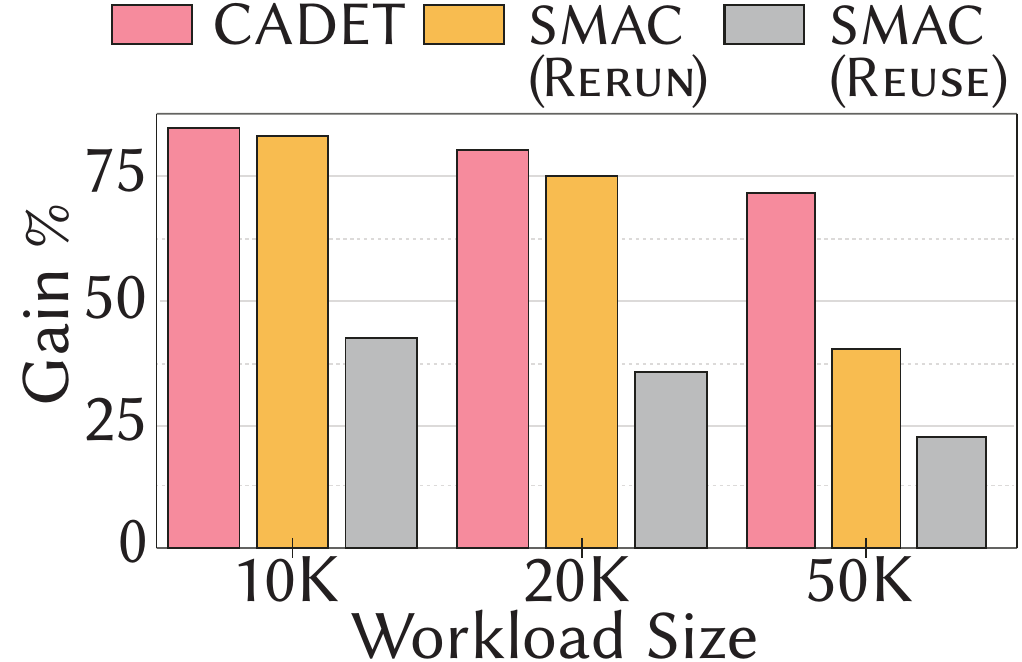}
        \label{fig:rq3_1_d}
        }\quad 
        \subfloat[Wallclock time.]{
            \includegraphics[width=0.4\linewidth]{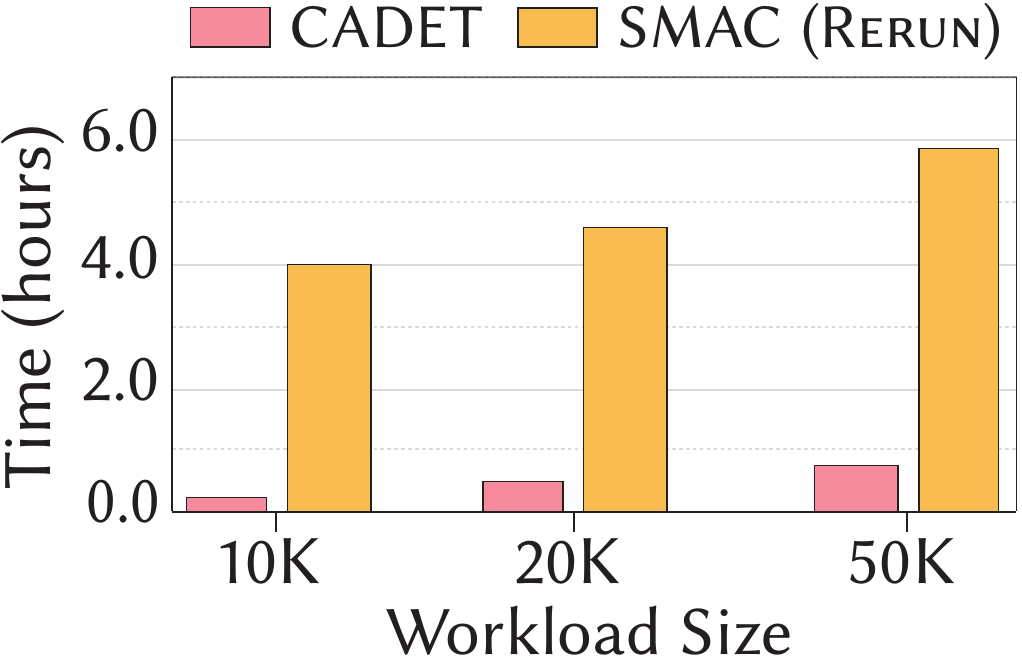}
            \label{fig:rq3_1_e}
    }
    \caption{\small{\tool vs. optimization with SMAC and PESMO.}}
    \label{tab:rq3_1}
\end{figure}

%
%
\noindent\textbf{RQ2-A.~How effective is optimization?~} In this RQ, we compare \tool with SMAC for fixing latency faults in \txtwo (for Image recognition). In general, the fixes recommended by \tool were observed to have better latency gains than the near-optimal configuration discovered by SMAC (see~\fig{rq3_1_a}). 
Most notably, in numerous cases, SMAC generated ``near-optimal'' configurations which in fact lead to significant deterioration in other non-functional properties such as energy consumption and heat dissipation (denoted by  $\blacklozenge$ in \fig{rq3_1_a}). Out of 21 latency faults in image recognition on \txtwo, we found 7 cases (33\%) where the near-optimal configuration caused a significant deterioration in energy consumption. In one case the optimal configuration increased the energy consumption by $96\%$! 
To overcome the above problem, we use a multi-objective optimization such as PESMO~\cite{hernandez2016predictive} to find a configuration that optimizes for energy and latency simultaneously. \tool can natively accommodate such multi-objective by slightly altering the counterfactual query (c.f.~\tion{root_cause}).

\fig{rq3_1_b} compares \tool with PESMO for fixing energy and latency faults in \txtwo (for Image recognition). As with the previous instance, we find that \tool performs better than PESMO, \ie, \tool recommends fixes misconfigurations that result in better gain than the Pareto-optimal configuration found by PESMO. 

\noindent\textbf{RQ2-B.~What is the analysis time of optimization?~}~The analysis time of all approaches are reported in~\fig{rq3_1_c}. Optimization methods run until they converge, however it is a common practice to assign a maximum budget (we use 4 hours) to limit inference time and cost. \tool was found to have comparatively lesser inference time compared to the other methods. While SMAC takes slightly more than 3 hours to find a near-optimum in \txtwo (image recognition), \tool resolved the faults in 20 minutes on average ($9\times$ faster). Likewise, PESMO takes close 4 hours compared to 45 minutes with \tool ($5\times$ faster). 

Since \tool uses causal models to infer candidate fixes using only the available observational data, it tends to be much faster than SMAC and PESMO. During incremental learning, \tool judiciously evaluates the most promising fixes until the fault is resolved. This limits the number of times the system is reconfigured and profiled. In contrast, optimization techniques sample the configuration space exhaustively to find an optimal configuration. Every sampled configuration is deployed and profiled. Optimization techniques sample several sub-optimal configurations in their search process and they are all evaluated. This issue is compounded when performing multi-objective optimization.

\noindent\textbf{RQ2-C.~How sensitive are optimization techniques to external changes in the software?~}~Configuration optimization methods are prone to be very sensitive to minor changes to the software stack. To demonstrate this, we use three larger additional image recognition workloads: 10K, 20K, and 50K test images (previous experiments used 5K test images). We evaluate two variants of SMAC: 
(1)~SMAC (\textsc{Reuse}) where we \textit{reuse} the near-optimum found with a 5K tests image on the larger workloads; and 
(2)~SMAC (\textsc{Rerun}) where we \textit{rerun} SMAC afresh on each workload.

Our experimental results demonstrate that \tool performs better than the two variants of SMAC (c.f. \fig{rq3_1_d}). SMAC (\textsc{Reuse}) performs the worst when the workload changes. With 10K images, reusing the near-optimal configuration from 5K images results in latency gain of 39\%, compared to 79\% with \tool. With a workload size of 50K images, SMAC (\textsc{Reuse}) only achieves a latency improvement of 24\% over the faulty configurations whereas \tool finds a fix that improves latency by $72\%$. SMAC (\textsc{Rerun}) performs better than SMAC (\textsc{Reuse}) but worse than \tool.  

In \fig{rq3_1_e} we report the inference times for SMAC (\textsc{Rerun}) and \tool. In keeping with the previous results, SMAC (\textsc{Rerun}) takes significantly longer than \tool, \ie, SMAC (\textsc{Rerun}) exceeds the 4 hour budget in all workloads whereas \tool takes at most 30 minutes for the largest workload to diagnose and fix the latency faults. We have to rerun SMAC every time the workload changes, and this limits its practical usability. In contrast, \tool incrementally updates the internal causal model with new samples from the larger workload to learn new relationships from the larger workload. Therefore, it is less sensitive and much faster. 

\subsection*{RQ3. Scalability of \tool}

\subsubsection*{\textbf{Motivation.}} To examine the scalability of \tool, we attempt to remedy latency issues in SQLite database deployed on \xavier by running the \texttt{speedtest} workload. SQLite was chosen because it offers a large number of configurable options, much more than  DNN-based applications. These configurations control the compute, memory, data reuse operations of SQLite. Further, each of these options can take on a large number of permitted values making SQLite a useful candidate to study scalability of \tool.

\subsubsection*{\textbf{Approach.}}~The causal graphs of previous experiments used 47 variables with 28 hardware, kernel/OS, and software configuration options in addition to 19 system events (c.f.~\tion{jetson_faults}). With SQLite, there are two additional scenarios: (a) 28 configuration options and 19 system events when we manually select the most relevant software/hardware options and events, (b) 242 configuration options and 88 events when we select all modifiable software and hardware options and system events, and (c) 242 configuration options and 288 events when we select not only all modifiable software and hardware options and system events but also intermediate {\tt tracepoint events}. For each scenario, we track the time required to discover the causal structure, evaluate the counterfactual queries, and total time to resolve latency faults. 

\subsubsection*{\textbf{Results.}}~Our findings, tabulated in \tab{scalability}, show that, with a larger number of configuration options, there are significantly more paths and queries from \tool. This adds to the causal graph discovery and query evaluation time. However, with as much as 242 configuration options and 88 events (\tab{scalability}, row 2), causal graph discovery takes roughly one minute, evaluating all 3136 queries takes roughly two minutes, and the total time to diagnose and fix a fault is roughly 28 minutes. This trend is observed even when the with 242 configuration options, 288 events (\tab{scalability}, row 3), and finer granularity of configuration values---the time required to causal model recovery is less than two minutes and the total time to diagnose and fix a fault is less than 2 hours. This indicates that \tool can scale to a much larger configuration space without an exponential increase in runtime for any of the intermediate stages. This property can be attributed to the sparsity of the causal graph (average degree of a node in \tab{scalability} is at most 3.6 and it reduces to 1.6 when the number of configurations increase). This makes sense because not all variables (\ie, configuration options and/or system events) affect non-functional properties and a high number of variables in the graph end up as isolated nodes. Therefore, the number of paths and consequently the evaluation time do not grow exponentially as the number of variables increase.


Finally, the latency gain associated with repairs from larger configuration space with configurations was similar to the original space of 28 configurations (both were 93\%). This indicates that: (a)~imparting domain expertise to select most important configuration options can speedup the inference time of \tool, and (b)~if the user chooses instead to use more configuration options (perhaps to avoid initial feature engineering), \tool can still diagnose and fix faults satisfactorily within a reasonable time. 

%


\begin{table}[t!]
\caption{Scalability of \tool for SQLite on \xavier.}
\label{tab:scalability}
\resizebox{\linewidth}{!}{
\begin{tabular}{V{2}rrrrrrV{2}rrrV{2}}
\clineB{7-9}{2}
\multicolumn{1}{l}{}               & \multicolumn{1}{l}{}           &
\multicolumn{1}{l}{}           &
\multicolumn{1}{l}{}           &
\multicolumn{1}{l}{}         & \multicolumn{1}{cV{2}}{}              & \multicolumn{3}{cV{2}}{Time/Fault (in sec.)}                                                    \bigstrut\\ \clineB{1-6}{2}
\multicolumn{1}{V{2}r}{\rotatebox{45}{Configs}} &
{\rotatebox{45}{Events}} &\multicolumn{1}{r}{\rotatebox{45}{Paths}} & \multicolumn{1}{r}{\rotatebox{45}{Queries}} &
\multicolumn{1}{r}{\rotatebox{45}{Degree}} &
\multicolumn{1}{rV{2}}{\rotatebox{45}{Gain (\%)}} & \multicolumn{1}{r}{\rotatebox{45}{Discovery}} & \multicolumn{1}{r}{\rotatebox{45}{Query Eval}} & \multicolumn{1}{rV{2}}{\rotatebox{45}{\bfseries Total}} \bigstrut\\ \hlineB{2}

\multicolumn{1}{c}{}& \multicolumn{1}{c}{}& \multicolumn{1}{c}{}& \multicolumn{1}{c}{}& \multicolumn{1}{c}{}& \multicolumn{1}{c}{}& \multicolumn{1}{c}{}\bigstrut\\[-1.5em] \hlineB{2}

\multicolumn{1}{V{2}r}{28}   & \multicolumn{1}{r}{19}                    & \multicolumn{1}{r}{32}        & \multicolumn{1}{r}{191} &   \multicolumn{1}{r}{3.6} & 93                              & 9                            & 14                            & \cellcolor{blue!10}\textbf{291}                     \bigstrut\\ 
\multicolumn{1}{V{2}r}{242}    & \multicolumn{1}{r}{88}                   & \multicolumn{1}{r}{181}       & \multicolumn{1}{r}{3136}   & \multicolumn{1}{r}{1.8}& 94                               & 68                           & 141                           & \cellcolor{blue!10}\textbf{1692}                    \bigstrut\\
\multicolumn{1}{V{2}r}{242}& \multicolumn{1}{r}{288}                      & \multicolumn{1}{r}{441}       & \multicolumn{1}{r}{22372} & \multicolumn{1}{r}{1.6}    & 92                             & 111                         & 854                             & \cellcolor{blue!10}\textbf{5312}           \bigstrut\\ \hlineB{2}

\end{tabular}}\vspace{-1em}

\end{table}

%% file: tables/rq1.tex
\begin{table*}[]
    \centering
    \caption{\small  Efficiency of \tool compared to other approaches. Cells highlighted in \colorbox{blue!10}{\bfseries blue} indicate improvement over faults.
    }
\vspace{-0.85em}
\subfloat[Single objective performance fault in latency and energy consumption.]{\scriptsize
    \label{tab:rq1_1}
    \resizebox{\textwidth}{!}{
    \begin{tabular}{@{}l|l|l|lllll|lllll|lllll|lllll|ll|}
        \clineB{4-25}{2}
        \multicolumn{1}{c}{}&\multicolumn{1}{c}{}  &  & \multicolumn{5}{c|}{Accuracy} & \multicolumn{5}{c|}{Precision} & \multicolumn{5}{c|}{Recall} & \multicolumn{5}{c|}{Gain} & \multicolumn{2}{c|}{Time$^\dagger$} \bigstrut\\ \clineB{4-25}{2}
        \multicolumn{1}{c}{}& \multicolumn{1}{c}{} &  & \multicolumn{1}{c}{\rotatebox{90}{\bfseries \tool}} &
        \multicolumn{1}{c}{\rotatebox{90}{\cbi}} & \multicolumn{1}{c}{\rotatebox{90}{\dd}} & \multicolumn{1}{c}{\rotatebox{90}{\encore}} & \multicolumn{1}{c|}{\rotatebox{90}{\bugdoc~}} & \multicolumn{1}{c}{\rotatebox{90}{\bfseries \tool}} &  
        \multicolumn{1}{c}{\rotatebox{90}{\cbi}} & \multicolumn{1}{c}{\rotatebox{90}{\dd}} & \multicolumn{1}{c}{\rotatebox{90}{\encore}} & \multicolumn{1}{c|}{\rotatebox{90}{\bugdoc~}} & \multicolumn{1}{c}{\rotatebox{90}{\bfseries \tool}} &
        \multicolumn{1}{c}{\rotatebox{90}{\cbi}} & \multicolumn{1}{c}{\rotatebox{90}{\dd}} & \multicolumn{1}{c}{\rotatebox{90}{\encore}} & \multicolumn{1}{c|}{\rotatebox{90}{\bugdoc~}} & \multicolumn{1}{c}{\rotatebox{90}{\bfseries \tool}} & \multicolumn{1}{c}{\rotatebox{90}{\cbi}} & \multicolumn{1}{c}{\rotatebox{90}{\dd}} & \multicolumn{1}{c}{\rotatebox{90}{\encore}} & \multicolumn{1}{c|}{\rotatebox{90}{\bugdoc
        ~}} & \multicolumn{1}{c}{\rotatebox{90}{\bfseries \tool}} &  \multicolumn{1}{c|}{\rotatebox{90}{Others}} \bigstrut[t]
        \\ \clineB{4-25}{2}
    \multicolumn{1}{c}{}&\multicolumn{1}{c}{}  & \multicolumn{1}{c}{} & \multicolumn{1}{c}{} & \multicolumn{1}{c}{} & \multicolumn{1}{c}{} & \multicolumn{1}{c}{} & \multicolumn{1}{c}{} \bigstrut\\[-1.4em]\hlineB{2}
     &  & Image & \cellcolor{blue!10}\bfseries84 & 66 & 65 & 68 & 71 & \cellcolor{blue!10}\bfseries 86 & 67 & 61 & 63 & 67 & \cellcolor{blue!10}\bfseries83  & 64 & 68 & 69 & 62 & \cellcolor{blue!10}\bfseries82  & 48 & 42 & 57 & 59 & \cellcolor{blue!10}\bfseries0.6 &4 \\
     &  & NLP & \cellcolor{blue!10}\bfseries75 &65 & 60 & 66 & 66 & \cellcolor{blue!10}\bfseries76  & 57 & 55 & 61 & 73 & \cellcolor{blue!10}\bfseries 71 & 74 & 68 & 67 & 65 & \cellcolor{blue!10}\bfseries74  & 54 & 59 & 62 & 58 & \cellcolor{blue!10}\bfseries0.4 & 4 \\
     &  & Speech & \cellcolor{blue!10}\bfseries76  & 64 & 63 & 63 & 72 & \cellcolor{blue!10}\bfseries76  & 58 & 69 & 61 & 71 & \cellcolor{blue!10}\bfseries81  & 73 & 61 & 63 & 69 & \cellcolor{blue!10}\bfseries76 & 59 & 53 & 55 & 66 & \cellcolor{blue!10}\bfseries0.7  & 4 \\
     \multirow{-4}{*}{\rotatebox{90}{\txtwo}} & \multirow{-4}{*}{\rotatebox{90}{Latency}} & x264 & \cellcolor{blue!10}\bfseries81  & 67 & 60 & 61 & 70 & \cellcolor{blue!10}\bfseries82  &69 & 58 & 65 & 66 & \cellcolor{blue!10}\bfseries78 & 64 & 67 & 63 & 72 & \cellcolor{blue!10}\bfseries85  & 69 & 72 & 68 & 71 & \cellcolor{blue!10}\bfseries1.4  & 4 \\ \hlineB{2}
    \multicolumn{1}{c}{}&\multicolumn{1}{c}{}  & \multicolumn{1}{c}{} & \multicolumn{1}{c}{} & \multicolumn{1}{c}{} & \multicolumn{1}{c}{} & \multicolumn{1}{c}{} & \multicolumn{1}{c}{} \bigstrut\\[-1.55em]\hlineB{2}
    
     &  & Image & \cellcolor{blue!10}\bfseries77 & 63 & 55 & 63 & 64 & \cellcolor{blue!10}\bfseries78 &56 & 58 & 66 & 65 & \cellcolor{blue!10}\bfseries80 & 69 & 55 & 63 & 68 & \cellcolor{blue!10}\bfseries83 & 59 & 50 & 35 & 51 & \cellcolor{blue!10}\bfseries0.4 & 4 \\
     &  & NLP & \cellcolor{blue!10}\bfseries76& 60 & 63 & 66 & 64 & \cellcolor{blue!10}\bfseries70  & 62 & 64 & 64 & 65 & \cellcolor{blue!10}\bfseries79  & 61 & 54 & 63 & 66 & \cellcolor{blue!10}\bfseries62  & 49 & 36 & 49 & 53 & \cellcolor{blue!10}\bfseries0.5  & 4 \\
     &  & Speech & \cellcolor{blue!10}\bfseries76  &66 & 65 & 61 & 71 & \cellcolor{blue!10}\bfseries75 & 55 & 59 & 54 & 68 & \cellcolor{blue!10}\bfseries78  &53 & 52 & 59 & 71 & \cellcolor{blue!10}\bfseries78 & 64 & 48 & 65 & 63 & \cellcolor{blue!10}\bfseries1.2  & 4 \\
     \multirow{-4}{*}{\rotatebox{90}{\xavier}} & \multirow{-4}{*}{\rotatebox{90}{Energy}}& x264 & \cellcolor{blue!10}\bfseries78  &62 & 57 & 59 & 67 & \cellcolor{blue!10}\bfseries83  & 63 & 53 & 61 & 66 & \cellcolor{blue!10}\bfseries78  & 67 & 53 & 54 & 72 & \cellcolor{blue!10}\bfseries 87  & 73 & 71 & 76 & 76 & \cellcolor{blue!10}\bfseries0.3  &4 \\ \hlineB{2}
    \end{tabular}
    }}
    \\
    \subfloat[Multi-objective non-functional faults in \textit{Energy, Latency}.]{
        \scriptsize
        \label{tab:rq2}
        \resizebox{\textwidth}{!}{
            \begin{tabular}{@{}r@{}ll|llll|llll|llll|llll|llll|ll|}
            \clineB{4-25}{2}
            &  &  & \multicolumn{4}{c|}{Accuracy} & \multicolumn{4}{c|}{Precision} & \multicolumn{4}{c|}{Recall} & \multicolumn{4}{c|}{Gain (Latency)} & \multicolumn{4}{c|}{Gain (Energy)}  & \multicolumn{2}{c|}{Time$^\dagger$} \bigstrut\\ \clineB{4-25}{2}
            
            &  &  & \rotatebox{90}{\bfseries \tool~} & \rotatebox{90}{\cbi} & \rotatebox{90}{\encore} & \rotatebox{90}{\bugdoc} & \rotatebox{90}{\bfseries \tool~}  & \rotatebox{90}{\cbi} & \rotatebox{90}{\encore} & \rotatebox{90}{\bugdoc} & \rotatebox{90}{\bfseries \tool~} & \rotatebox{90}{\cbi} & \rotatebox{90}{\encore} & \rotatebox{90}{\bugdoc} & \rotatebox{90}{\bfseries \tool~} & \rotatebox{90}{\cbi} & \rotatebox{90}{\encore} & \rotatebox{90}{\bugdoc} & \rotatebox{90}{\bfseries \tool~} & \rotatebox{90}{\cbi} & \rotatebox{90}{\encore} & \rotatebox{90}{\bugdoc} & \rotatebox{90}{\bfseries \tool~}  & \rotatebox{90}{Others} \\ \clineB{4-25}{2}
            
            \multicolumn{1}{l}{}&\multicolumn{1}{l}{}  & \multicolumn{1}{l}{} & \multicolumn{1}{l}{} & \multicolumn{1}{l}{} & \multicolumn{1}{l}{} & \multicolumn{1}{l}{} & \multicolumn{1}{l}{} & \multicolumn{1}{l}{} & \multicolumn{1}{l}{} \\[-0.85em]\hlineB{2}

            & \multicolumn{1}{l|}{} & Image & \cellcolor{blue!10}\textbf{80} & 54 & 55 & 65 & \cellcolor{blue!10}\textbf{77} & 53 & 54 & 62 & \cellcolor{blue!10}\textbf{81} & 59 & 59 & 62 & \cellcolor{blue!10}\textbf{84} & 53 & 61 & 65 & \cellcolor{blue!10}\textbf{75} & 38 & 46 & 44 & \cellcolor{blue!10}\textbf{0.9} & 4 \\
            
            & \multicolumn{1}{l|}{} & NLP & \cellcolor{blue!10}\textbf{76} &51 & 56 & 65 & \cellcolor{blue!10}\textbf{77} & 42 & 56 & 63 & \cellcolor{blue!10}\textbf{79} & 59 & 62 & 65 & \cellcolor{blue!10}\textbf{84} & 53 & 59 & 61 & \cellcolor{blue!10}\textbf{67}  & 41 & 27 & 48 & \cellcolor{blue!10}\textbf{0.5}  & 4 \\
            
            & \multicolumn{1}{l|}{} & Speech & \cellcolor{blue!10}\textbf{81}  & 50 & 61 & 65 & \cellcolor{blue!10}\textbf{80} & 44 & 53 & 62 & \cellcolor{blue!10}\textbf{81}  & 51 & 59 & 64 & \cellcolor{blue!10}\textbf{88}  & 55 & 55 & 62 & \cellcolor{blue!10}\textbf{77}  & 43 & 43 & 41  & \cellcolor{blue!10}\textbf{1.1}  & 4 \\
            \multirow{-4}{*}{\rotatebox{90}{Energy +}} & \multicolumn{1}{l|}{\multirow{-4}{*}{\rotatebox{90}{Latency}}} & \multicolumn{1}{l|}{x264} & \cellcolor{blue!10}\textbf{81}  & 54 & 55 & 66 & \cellcolor{blue!10}\textbf{83}  & 50 & 54 & 67 & \cellcolor{blue!10}\textbf{80}  & 63 & 62 & 61 & \cellcolor{blue!10}\textbf{75}  & 62 & 64 & 66 & \cellcolor{blue!10}\textbf{76}  & 64 & 66 & 64  & \cellcolor{blue!10}\textbf{1}  & 4 \\\hlineB{2}
           \multicolumn{10}{l}{$^\dagger$ Wallclock time in hours}\bigstrut
           \end{tabular}
    }}\vspace{-2.5em}
\end{table*}

%% file: 9_related.tex
\section{Related Work}
\label{sect:related}
\noindent

\noindent\textbf{Performance Faults in Configurable Systems.} Previous empirical studies have shown that a majority of performance issues are due to misconfigurations~\cite{han2016empirical}, with severe consequences in production environments~\cite{tang2015holistic,maurer2015fail}, and configuration options that cause such performance faults force the users to tune the systems themselves~\cite{zhang2021evolutionary}. However, the configuration options keep increasing over time~\cite{xu2015hey} and make the configuration-aware testing a difficult endeavor, where testing can be only practical by evaluating a representative sample of all possible configurations~\cite{qu2008configuration}, by employing sampling strategies to discover potential (performance) faults~\cite{medeiros2016comparison}. Previous works have used static and dynamic program analysis to identify the influence of configuration options on performance~\cite{velez2019configcrusher,velez2021white,li2020statically} and to detect and diagnose misconfigurations~\cite{XJHZLJP:OSDI16,attariyan2010automating,zhang2013automated,attariyan2012x}. 
None of the white-box analysis approaches target configuration space across system stack, where it limits their applicabilities in identifying the true causes of a performance fault. The evaluation results indicate that by targeting the whole system stack, \tool goes beyond the existing approaches in terms of detecting and fixing performance faults.


\noindent\textbf{Performance Modeling and Optimization.}
Performance behavior of configurable systems is complex (non-linear, non-convex, and multi-modal)~\cite{wang2018understanding,jamshidi2016uncertainty}, where such behavior becomes even more intricate when multiple objectives are needed to be traded-off for performance-related tasks~\cite{kolesnikov2019tradeoffs,nardi2019hypermapper,iqbal2020flexibo}. A common strategy to understand such complex performance behavior is to use machine-learning by building a black box model that characterizes system performance~\cite{siegmund2015performance,valov2017transferring,guo2013variability}. The learned performance model can then be used for performance debugging ~\cite{siegmund2015performance,guo2013variability,wang2018understanding} and tuning~\cite{H:AS}.
While the sheer size of the configuration space and complex performance behavior prohibit any guarantee of finding a globally optimal configuration, optimization algorithms attempt to find near-optimal configurations under a limited sampling budget using a clever combination of sampling, model construction, and search. While there is no silver bullet~\cite{grebhahn2019predicting}, several methods have been attempted including hill climbing~\cite{XLRXZ:WWW04}, optimization via guessing~\cite{OK:SIGMETRICS07}, Bayesian optimization~\cite{JC:MASCOTS16}, and multi-objective optimization~\cite{wu2015deep,iqbal2020flexibo}. 
However, if one of the contextual/environmental conditions (e.g., hardware, workload) of the system changes, one needs to rerun the optimization approach again~\cite{JVKS:FSE18}, therefore, transfer learning in terms of transferring measurement data or extracting some types of knowledge across environments has been used~\cite{valov2017transferring,JSVKPA:ASE17,JVKSK:SEAMS17,javidian2019transfer,iqbal2019transfer,krishna2019whence,valov2020transferring}. For a more comprehensive treatment of the literature, we refer to~\cite{pereira2019learning,siegmund2020dimensions}.
Although black-box models have used extensively for optimization, they cannot diagnose non-functional faults as they are unaware of underlying causal factors due to their blind exploration without having access to the underlying causal relationships.

\noindent\textbf{Statistical and Model-based Debugging.}
Debugging approaches such as \textsc{Statistical Debugging}~\cite{song2014statistical}, 
\textsc{HOLMES}~\cite{chilimbi2009holmes}, \textsc{XTREE}~\cite{krishna2017less}, \bugdoc~\cite{lourencco2020bugdoc}, \encore~\cite{lourencco2020bugdoc}, \textsc{Rex}~\cite{mehta2020rex}, and \textsc{PerfLearner}~\cite{han2018perflearner} have been proposed to detect root causes of system faults. These methods make use of statistical diagnosis and pattern mining to rank the probable causes based on their likelihood of being the root causes of faults. However, these approaches may produce correlated predicates that lead to incorrect explanations.



\noindent\textbf{Causal Testing and Profiling.} 
Causal inference has been used for fault localization \cite{baah2010causal,feyzi2019inforence}, resource allocations in cloud systems~ \cite{geiger2016causal}, and causal effect estimation for advertisement recommendation systems \cite{bottou2013counterfactual}. More recently, \textsc{AID}~\cite{fariha2020causality} detects root causes of a intermittent software failure using fault injection as interventions and utilizing a guided search to pinpoint the root causes and generate an explanation how the root cause triggers the failure. Similarly \textsc{LDX}~\cite{kwon2016ldx} determines causal dependencies of events for detection of information leak and security attacks. \textsc{Causal Testing} and \textsc{Holmes}~\cite{johnson2020causal} modifies the system inputs to observe behavioral changes in order to build a causal model  and then utilizes counterfactual reasoning to find the root causes of bugs. Causal profiling approaches like \textsc{CoZ}~\cite{curtsinger2015coz} points developers where optimizations will improve performance and quantifies their potential impact, where the interventions happen by virtual program speedups. Causal inference methods like \textsc{X-Ray}~\cite{attariyan2012x} and \textsc{ConfAid}~\cite{attariyan2010automating} had previously been applied to analyze program failures using run-time control and data flow behavior. All approaches above are either orthogonal or complimentary to \tool, mostly they focus on functional bugs (e.g., \textsc{Causal Testing}) or if they are performance related, they are not configuration-aware (e.g., \textsc{CoZ}).

%% file: 10_conclusion.tex
\section{Conclusion}
\label{sect:conclusion}

Modern computer systems are highly-configurable with thousands of interacting configurations with a complex performance behavior. Misconfigurations in these systems can elicit complex interactions between software and hardware configuration options resulting in non-functional faults. We propose \tool, a novel approach for diagnostics that learns and exploits the causal structure of configuration options, system events, and performance metrics. Our evaluation shows that \tool effectively and quickly diagnoses the root cause of non-functional faults and recommends high-quality repairs to mitigate these faults.  

\section*{Acknowledgements} We like to thank Christian K$\ddot{\text{a}} $stner, Sven Apel, Yuriy Brun, Emery Berger, Tianyin Xu, Vivek Nair, Jianhai Su, Miguel Velez, Tobius D$\ddot{\text{u}}$rschmid for their valuable feedback and suggestions in improving the paper. This work was partially supported by NASA (RASPBERRY-SI Grant Number 80NSSC20K1720) and NSF (SmartSight Award 2007202). 